\documentclass[%
reprint,
superscriptaddress,
nofootinbib,
amsmath,amssymb,
aps,
]{revtex4-2}

\usepackage{graphicx}% Include figure files
\usepackage{dcolumn}% Align table columns on decimal point
\usepackage{bm}% bold math
\usepackage{braket}
\usepackage{upgreek}
\usepackage{xcolor}
\usepackage{afterpage}
\usepackage{relsize}
\usepackage{multirow}
\usepackage{float}
\usepackage{physics}
\usepackage{makecell}
\usepackage{placeins}
\usepackage{amsthm}
\usepackage{soul}
\usepackage[colorlinks=true,urlcolor=blueprl,citecolor=blueprl,linkcolor=blueprl]{hyperref}
\usepackage{lineno}
\usepackage[symbol]{footmisc}
\usepackage{hhline}
\usepackage[normalem]{ulem}
\usepackage{booktabs}
\usepackage{array}
\usepackage{nicefrac}

\newcommand{\ozlem}[1]{\textcolor{black}{#1}}
\newcommand{\AD}[1]{\textcolor{black}{#1}}

\newcolumntype{P}[1]{>{\centering\arraybackslash}p{#1}}

\usepackage{hyperref}
\let\svthefootnote\thefootnote
\newcommand\freefootnote[1]{%
  \let\thefootnote\relax%
  \footnotetext{#1}%
  \let\thefootnote\svthefootnote%
}

\definecolor{blueprl}{RGB}{46,48,146}

\def\UQA{Centre of Excellence for Quantum Computation and Communication Technology, School of Mathematics and Physics, University of Queensland, St Lucia, QLD 4072, Australia}

\def\UTS{Centre for Quantum Software and Information, University of Technology Sydney, Sydney,
New South Wales 2007, Australia}

\begin{document}

%\preprint{APS/123-QED}

%\title{A unified low-squeezing optical platform for fault-tolerant Gottesman-Kitaev-Preskill states and other non-Gaussian states}
\title{Near-Optimal Bell Nonlocality with GKP and Entangled Cat States}

\author{\"{O}zlem Erk{\i}l{\i}\c{c}$^{*}$}
\freefootnote{$^*$ \href{ozlemerkilic1995@gmail.com}{ozlemerkilic1995@gmail.com}}
%\freefootnote{$^\dagger$ These authors contributed equally}
\affiliation{\UTS}

\author{Aritra Das}
\affiliation{\UTS}

\author{S. Nibedita Swain}
%\affiliation{\UTSMaths}
\affiliation{\UTS}
\affiliation{\UQA}

\author{Simon Devitt}
\affiliation{\UTS}

\author{Timothy C. Ralph}
\affiliation{\UQA}

\date{\today}

\begin{abstract}
    \ozlem{Gottesman-Kitaev-Preskill~(GKP) states are widely studied as a bosonic encoding for fault-tolerant quantum computing because small displacement errors can be identified and corrected through syndrome measurements. However, fault-tolerant operation requires substantially greater GKP squeezing than is currently available in optical platforms. It is therefore important to identify quantum-information tasks that can be realised with finite-squeezing GKP states. Here, we study Bell nonlocality in finite-energy GKP Bell states under experimentally motivated measurement constraints. Although finite-energy GKP Bell states support Bell nonlocality even at low squeezing, logical Pauli measurements alone cannot reveal it. Accessing the maximum CHSH violation supported by the state generally requires logical non-Clifford rotations. We instead show that the required non-Gaussian resource can reside in the measurement itself. Photon-number-resolving detection combined with Gaussian preprocessing and adaptive feed-forward closely approaches the maximum violation of the CHSH Bell inequality without implementing logical non-Clifford gates. The same measurement principle strongly enhances Bell violations for cat-code Bell pairs and entangled coherent states, demonstrating that its usefulness extends beyond the GKP lattice. Our results identify non-Gaussian measurements as a practical route to revealing Bell nonlocality in finite-energy bosonic states.}
\end{abstract}

\maketitle

\section{\label{sec:introduction}Introduction}
Bell nonlocality captures a fundamental departure of quantum mechanics from classical descriptions of nature, demonstrating correlations that cannot be explained by local hidden-variable theories~\cite{bell1964einstein,brunner2014bell}. Following landmark experimental demonstrations of Bell-inequality violations~\cite{freedman1972experimental,aspect1981experimental,aspect1982experimental,weihs1998violation}, Bell nonlocality has emerged as an important resource for device-independent quantum information processing~\cite{brunner2014bell}. In particular, violations of the Clauser-Horne-Shimony-Holt~(CHSH) inequality~\cite{clauser1969proposed} enable self-testing, where quantum states and measurements are certified from observed correlations~\cite{kaniewski2016analytic, zhang2018experimentally, vsupic2020self}; device-independent quantum key distribution (DI-QKD), where security does not rely on detailed device models~\cite{acin2006bell, acin2007device, vazirani2019fully, zhang2022device, nadlinger2022experimental, zapatero2023advances}; and certified randomness generation, where Bell violations certify genuine randomness~\cite{pironio2010random,acin2016certified,liu2018device,shalm2021device}. Realising these applications, however, depends critically on the available local measurements, as the optimal measurements for revealing nonlocal correlations can be experimentally demanding. Bridging the gap between optimal and physically accessible measurements therefore remains an important challenge.

This measurement challenge is particularly relevant in continuous-variable (CV) systems, where information is encoded in continuous degrees of freedom such as the amplitude and phase quadratures of an optical field~\cite{weedbrook2012gaussian}. CV states can be categorised into Gaussian states, which are fully characterised by their first and second statistical moments, and non-Gaussian states, whose description generally requires higher-order moments. Although Gaussian states can be entangled, their positive Wigner functions together with Gaussian measurements admit a local hidden-variable description and therefore cannot produce Bell violations within an entirely Gaussian setting~\cite{braunstein2005quantum}. Bell nonlocality in CV systems thus requires a non-Gaussian resource, either in the state or the measurement~\cite{braunstein2005quantum}. Early approaches introduced such resources through non-Gaussian measurements, including pseudospin observables~\cite{chen2002maximal} and displaced photon-number parity measurements~\cite{banaszek1998nonlocality,banaszek1999testing}. In parallel, non-Gaussian states themselves provided another route to CV Bell nonlocality. Entangled coherent states~(ECSs), introduced as a setting for Bell tests of nonorthogonal states~\cite{sanders1992entangled, mann1995bell}, have since become among the most extensively studied examples, with measurement strategies ranging from homodyne detection combined with nonlinear local rotations~\cite{stobinska2007violation} to displaced photon-number parity measurements~\cite{wilson2002quantum, jeong2003quantum} and pseudospin observables~\cite{de2023entangled}. While these approaches can yield strong CHSH violations, their performance and experimental requirements depend strongly on the measurement scheme.

Another important class of non-Gaussian CV states is Gottesman-Kitaev-Preskill (GKP) states~\cite{gottesman2001encoding}, which encode discrete-variable quantum information into the continuous quadratures of a bosonic mode. Originally introduced for bosonic quantum error correction, GKP states are a key resource for fault-tolerant CV quantum computation, with optical generation schemes based on cat-state breeding~\cite{vasconcelos2010all,weigand2018generating,konno2024logical,winnel2024deterministic,erkilicc2026unified} and Gaussian boson sampling~\cite{zhong2019experimental,takase2023gottesman,deng2023gaussian,liu2025robust,larsen2025integrated}. \ozlem{Despite their potential, GKP states remain challenging to realise experimentally. Fault-tolerant operation typically requires approximately $9.75\,\mathrm{dB}$ of GKP squeezing~\cite{larsen2025integrated}, while optical demonstrations have so far reached around $2.5\,\mathrm{dB}$~\cite{konno2024logical,larsen2025integrated}. This gap highlights the importance of identifying quantum-information tasks that remain achievable with experimentally accessible GKP squeezing. In particular,} GKP Bell states encode maximally entangled qubits and can, in principle, attain the Tsirelson bound~\cite{cirel1980quantum}. Achieving these correlations with experimentally accessible measurements, however, remains non-trivial. Recent work has shown that homodyne detection combined with Gaussian operations and logical Pauli measurements is insufficient to violate the CHSH inequality with GKP Bell states~\cite{yang2026robust}. Existing proposals circumvent this restriction by implementing tilted logical measurements using non-Clifford operations~\cite{marshall2014device}. Such operations are considerably more demanding than the Gaussian and Clifford toolbox naturally available for GKP states~\cite{hastrup2021cubic}. \ozlem{In fault-tolerant GKP architectures, non-Clifford operations require gate teleportation with ancillary GKP-encoded magic states, whose preparation and distillation incur substantial resource overhead~\cite{baragiola2019all,yamasaki2020cost,konno2021non}.} This motivates a central question: are non-Clifford logical operations fundamentally necessary to access the Bell nonlocality of GKP states?

In this work, we characterise how measurement constraints determine the accessible Bell nonlocality of finite-energy GKP states. We derive CHSH bounds as a function of GKP squeezing under different restrictions on the local measurements, establishing the ultimate performance available to experimentally motivated measurement classes. We then show that violating CHSH does not require logical non-Clifford operations. Non-Gaussian photon-number-resolving detection~(PNRD) combined with Gaussian preprocessing yields strong violations approaching these bounds. Finally, we apply the same measurement framework to entangled coherent states, substantially reducing the coherent-state amplitude required for Bell violation and approaching the Tsirelson bound. These results establish a general measurement-based route to \ozlem{demonstrate} Bell nonlocality in bosonic quantum states without implementing logical non-Clifford rotations.

\section{Bell nonlocality of finite-energy GKP states}
In this section, we first quantify how much CHSH nonlocality can be \ozlem{obtained} from finite-energy GKP Bell states under different restrictions on the local measurements. By comparing standard periodic homodyne binning, optimised homodyne binning, and fully unrestricted measurements, we distinguish the intrinsic nonlocality of the state from that accessible to a given measurement toolbox.

\subsection{Finite-energy GKP Bell states}
\ozlem{Ideal GKP states consist of infinitely sharp peaks extending over an infinite phase-space lattice, and therefore require infinite squeezing and infinite energy. They are consequently not physically realisable.} Finite-energy GKP states instead approximate the ideal codewords using a lattice of finitely squeezed states modulated by a Gaussian envelope. We denote the resulting logical states by $\ket{\bar{0}_{\Delta}}$ and $\ket{\bar{1}_{\Delta}}$, where $\Delta$ characterises the finite squeezing.

To define these states, we first define a normalised Gaussian peak of width $\Delta$ centred at $q_0$ as
\begin{equation}
    \langle{q}\ket{G_{\Delta}(q_0)} = \frac{1}{(\pi\Delta^2)^{1/4}} \exp\left[
        -\frac{(q-q_0)^2}{2\Delta^2}
    \right].
    \label{eq:gaussian_envelope}
\end{equation}
The finite-energy logical GKP states can then be expressed as~\cite{gottesman2001encoding}
\begin{align}
    \label{eq:gkp_codewords}
    \ket{\bar{0}_{\Delta}}
    &=
    \frac{1}{\sqrt{\mathcal{N}_0}}
    \sum_{s\in\mathbb{Z}}
    e^{-2\pi\Delta^2s^2}
    \ket{G_{\Delta}(2s\sqrt{\pi})},
    \\ \nonumber
    \ket{\bar{1}_{\Delta}}
    &=
    \frac{1}{\sqrt{\mathcal{N}_1}}
    \sum_{s\in\mathbb{Z}}
    e^{-\frac{\pi\Delta^2}{2}(2s+1)^2}
    \ket{G_{\Delta}((2s+1)\sqrt{\pi})}.
\end{align}
where $\mathcal{N}_0$ and $\mathcal{N}_1$ are normalisation constants. From these codewords, we construct the finite-energy GKP Bell state
\begin{equation}
    \ket{\Phi_{\Delta}}=\frac{        \ket{\bar{0}_{\Delta},\bar{0}_{\Delta}}+
        \ket{\bar{1}_{\Delta},\bar{1}_{\Delta}}}{\sqrt{\mathcal{N}}},
    \label{eq:finite_gkp_bell}
\end{equation}
where $\mathcal{N}$ represents the normalisation constant. At finite squeezing, the logical codewords $\ket{\bar{0}_{\Delta}}$ and $\ket{\bar{1}_{\Delta}}$ are not perfectly orthogonal, and the resulting Bell state therefore differs from an ideal maximally entangled logical qubit pair. As the squeezing increases, corresponding to $\Delta\rightarrow 0$, the individual peaks become increasingly narrow, the overlap between the logical codewords vanishes, and $\ket{\Phi_{\Delta}}$ approaches the ideal GKP Bell state. Throughout this work, we express the finite squeezing in decibels\ozlem{~(dB)} as $s_{\mathrm{GKP}} = -10\log_{10}\left(\Delta^2\right)$.

\subsection{Measurement-constrained CHSH bounds}
To determine how measurement constraints limit the observable Bell nonlocality of finite-energy GKP states, we compare three levels of measurement capability: standard homodyne detection with periodic GKP binning, homodyne detection with optimised binary binning, and unrestricted dichotomic measurements. Here, dichotomic refers to a measurement with two outcome labels, assigned the values $+1$ and $-1$ in the CHSH test~\cite{brunner2014bell}. Comparing these benchmarks allows us to distinguish limitations imposed by the measurement strategy from those intrinsic to the finite-energy state.

For two dichotomic measurement settings $A_0$ and $A_1$ for Alice and $B_0$ and $B_1$ for Bob, the CHSH parameter is defined as
\begin{equation}
    S =
    \langle A_0 B_0\rangle
    + \langle A_0 B_1\rangle
    + \langle A_1 B_0\rangle
    - \langle A_1 B_1\rangle.
    \label{eq:chsh}
\end{equation}
Local hidden-variable theories are bounded by $|S|\leq 2$, whereas quantum mechanics allows values up to the Tsirelson bound $|S|\leq 2\sqrt{2}$.

\subsubsection{Periodic GKP binning}
We first consider the standard GKP measurement strategy based on homodyne detection with periodic binning. For an ideal GKP state, logical $Z$ and $X$ measurements can be implemented by homodyne detection of the $q$ and $p$ quadratures, respectively, followed by assigning each measurement outcome to the nearest logical lattice point. We define the corresponding periodic binning function as
\begin{equation}
    f_{\mathrm{GKP}}(x)
    =
    \begin{cases}
        +1,
        & x \in
        \left[
            \left(2k-\frac{1}{2}\right)\sqrt{\pi},
            \left(2k+\frac{1}{2}\right)\sqrt{\pi}
        \right), \\[6pt]
        -1,
        & x \in
        \left[
            \left(2k+\frac{1}{2}\right)\sqrt{\pi},
            \left(2k+\frac{3}{2}\right)\sqrt{\pi}
        \right),
    \end{cases}
    \label{eq:gkp_periodic_binning}
\end{equation}
where $k\in\mathbb{Z}$. The corresponding binned-homodyne observables are
\begin{align}
    \hat{Z}_{\mathrm{hom}}
    &=
    \int_{-\infty}^{\infty}
    dq\,f_{\mathrm{GKP}}(q)\ket{q}\!\bra{q},
    \nonumber\\
    \hat{X}_{\mathrm{hom}}
    &=
    \int_{-\infty}^{\infty}
    dp\,f_{\mathrm{GKP}}(p)\ket{p}\!\bra{p}.
    \label{eq:gkp_homodyne_observables}
\end{align}
For the numerical calculations, we truncate the oscillator Hilbert space to the first $N_{\mathrm{cut}}$ Fock states using $P_N=\sum_{n=0}^{N_{\mathrm{cut}}-1}\ket{n}\!\bra{n}$. The homodyne observables are defined in the continuous quadrature basis before projection \ozlem{onto} the truncated space. For an arbitrary binary binning function $f(x)\in\{-1,+1\}$, we define
\begin{align}
    A_q^{(N)}[f]
    &=
    \int_{-\infty}^{\infty}
    dq\,f(q)\,
    P_N\ket{q}\!\bra{q}P_N,
    \nonumber\\
    A_p^{(N)}[f]
    &=
    \int_{-\infty}^{\infty}
    dp\,f(p)\,
    P_N\ket{p}\!\bra{p}P_N.
    \label{eq:projected_homodyne}
\end{align}
Alice's observables for the periodic-binning strategy are therefore
\begin{equation}
    A_0=A_q^{(N)}[f_{\mathrm{GKP}}],
    \qquad
    A_1=A_p^{(N)}[f_{\mathrm{GKP}}].
    \label{eq:alice_periodic}
\end{equation}

At finite squeezing, the codewords
$\ket{\bar{0}_{\Delta}}$ and $\ket{\bar{1}_{\Delta}}$ are not exactly
orthogonal. We therefore construct an orthonormal basis
$\{\ket{0_L},\ket{1_L}\}$ for their two-dimensional span using symmetric
Löwdin orthogonalisation~\cite{lowdin1950non, torun2023coherence}, which preserves the symmetries of the codewords (refer to Appendix~\ref{app:measurement_optimisation} for a detailed explanation). Defining $V_L =\begin{pmatrix}\ket{0_L} & \ket{1_L}\end{pmatrix}$,
any operator $O$ acting in the truncated Fock space can be represented on
the exact support of the finite-energy GKP state as $O_L = V_L^\dagger O V_L$.

This is an exact change of basis on the state support and does not replace
the finite-energy GKP states by ideal qubits. For fixed Alice observables, Bob's two CHSH correlation operators are expressed as
\begin{align}
    C_0 &=
    \mathrm{Tr}_A\!\big[
        \rho_{AB}\big((A_0+A_1)\otimes I\big)\big],
    \nonumber\\
    C_1 &=
    \mathrm{Tr}_A\!\big[
        \rho_{AB}\big((A_0-A_1)\otimes I\big)
    \big].
    \label{eq:bob_correlation_operators}
\end{align}
Restricting these operators to the Löwdin basis gives $C_y^{(L)}=V_L^\dagger C_y V_L$. For each setting $y$, we define $U_y$ as the basis rotation that diagonalises the correlation operator according to $U_y^\dagger C_y^{(L)} U_y=\Lambda_y$ for $y\in\{0,1\}$, or equivalently, $C_y^{(L)}=U_y\Lambda_y U_y^\dagger$. The corresponding unrestricted binary observable is
\begin{equation}
    B_y^{\mathrm{opt}}
    =
    \operatorname{sign}\!\left(C_y^{(L)}\right)
    =
    U_y\operatorname{sign}(\Lambda_y)U_y^\dagger.
    \label{eq:bob_optimal}
\end{equation}
For the periodic-homodyne benchmark, Bob keeps this measurement direction but is restricted to homodyne detection with periodic GKP binning. Defining the support-restricted homodyne observable
\begin{equation}
    A_{q,L}^{(N)}[f]
    =
    V_L^\dagger A_q^{(N)}[f]V_L,
\end{equation}
Bob's periodic-binning benchmark observables are
\begin{equation}
    B_y^{\mathrm{per}}
    =
    U_y A_{q,L}^{(N)}[f_{\mathrm{GKP}}] U_y^\dagger.
    \label{eq:bob_periodic}
\end{equation}
\ozlem{Physically, $U_y$ denotes} an ideal logical basis rotation on the two-dimensional support of the finite-energy GKP state. This construction isolates the restriction imposed by periodic homodyne binning, without assuming a particular oscillator-level implementation of $U_y$. At finite squeezing, $B_y^{\mathrm{per}}$ generally differs from $B_y^{\mathrm{opt}}$, quantifying the limitation imposed by periodic homodyne binning.

\subsubsection{Optimised homodyne binning}
The periodic GKP binning rule is naturally suited to distinguishing the \ozlem{ideal} logical codewords, but is not necessarily optimal for maximising the CHSH parameter at finite squeezing. To determine the limitation imposed by the binning rule itself, we \ozlem{restrict the measurement to} homodyne detection while allowing each quadrature outcome to be assigned independently to $\pm1$. For Bob's measurement setting $y$, we consider an arbitrary binary binning function $f_y(q)\in\{-1,+1\}$, giving the homodyne observable
\begin{equation}
    B_y^{\mathrm{hom}}[f_y]
=
U_y A_{q,L}^{(N)}[f_y] U_y^\dagger.
    \label{eq:bob_optimised_homodyne}
\end{equation}
\ozlem{For fixed Alice observables, the contribution of Bob's homodyne outcome $q$ to the CHSH parameter is determined by}
\begin{equation}
    d_y(q)
=
\bra{q}
V_L U_y^\dagger C_y^{(L)} U_y V_L^\dagger
\ket{q},
    \label{eq:homodyne_score}
\end{equation}
where $C_y^{(L)}$ is the corresponding CHSH correlation operator defined above. The CHSH parameter is therefore maximised by assigning each outcome according to the sign of its contribution,
\begin{equation}
    f_y^{\mathrm{opt}}(q)
    =
    \operatorname{sign}\!\left[d_y(q)\right].
    \label{eq:optimal_homodyne_binning}
\end{equation}
Unlike periodic GKP binning, this assignment is determined directly by the
CHSH correlations \ozlem{in the physical state} rather than by proximity to the logical lattice points.

\subsubsection{Unrestricted measurements}
\begin{figure*}[htbp]
%\hspace*{-0.3cm}
\includegraphics[scale=0.46]{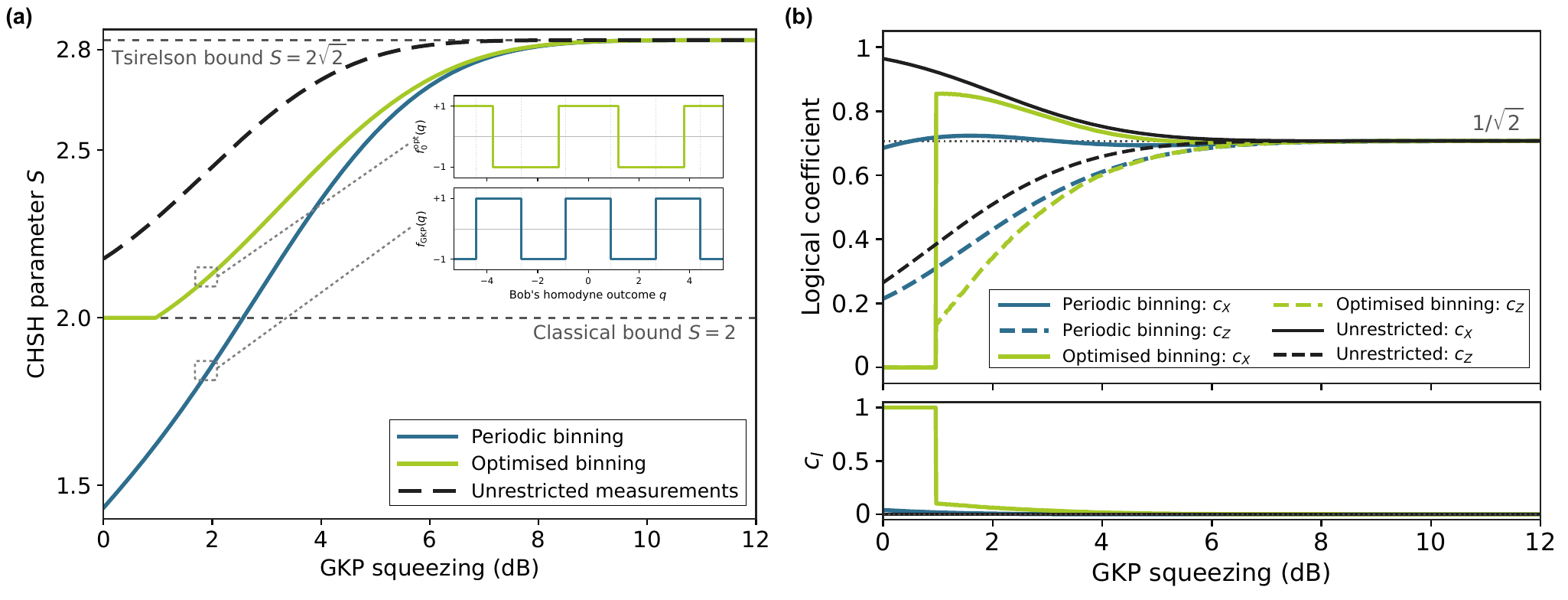}\hspace*{-0.1cm}
\caption{\label{fig:figure1}
Measurement-constrained Bell nonlocality of finite-energy GKP Bell states.
(a) Maximum CHSH parameter as a function of GKP squeezing for periodic GKP binning, optimised homodyne binning, and unrestricted dichotomic measurements (Refer to Appendix~\ref{app:parameters_used} for the simulation parameters). The horizontal lines indicate the classical bound $S=2$ and the Tsirelson bound $S=2\sqrt{2}$. \ozlem{The inset compares Bob's binary $q$-homodyne decision functions at $2\,\mathrm{dB}$: the upper panel shows the optimised assignment $f^{\mathrm{opt}}_0(q)$, while the lower panel shows the periodic GKP rule $f_{\mathrm{GKP}}(q)$. Whereas periodic binning alternates between equal lattice intervals, the optimised filter assigns outcomes according to their contribution to the CHSH parameter. The dotted guides indicate the corresponding $2\,\mathrm{dB}$ operating point in the main panel.} 
(b) Effective logical decomposition of Bob's $B_0$ observable, $B_0=c_I I+c_X X_L+c_Z Z_L$, for the corresponding measurement strategies. Solid and dashed curves show the $X_L$ and $Z_L$ coefficients, respectively, while the lower panel shows the identity contribution $c_I$. The unrestricted measurement approaches the ideal qubit direction $c_X=c_Z=1/\sqrt{2}$ with increasing squeezing.
}
\end{figure*}
%Finally, we remove all restrictions on the local measurements to determine the maximum CHSH violation supported by the finite-energy GKP Bell state. 
\ozlem{Finally, we remove the measurement restrictions considered in the previous sections to determine the maximum CHSH violation supported by finite-energy GKP Bell states under local bipartite measurement.} Unlike the preceding homodyne-based strategies, Alice and Bob are now allowed to perform arbitrary dichotomic measurements. The resulting value therefore provides a \ozlem{measurement setting-independent} upper benchmark for the CHSH correlations accessible at each squeezing level.

Since the finite-energy GKP Bell state has support on a two-dimensional subspace for each mode, it can be expressed in its Schmidt basis as
\begin{equation}
    \ket{\Phi_{\Delta}}
    =
    \lambda_0 \ket{0,0}
    +
    \lambda_1 \ket{1,1},
    \qquad
    \lambda_0^2+\lambda_1^2=1,
    \label{eq:gkp_schmidt}
\end{equation}
where $\lambda_0$ and $\lambda_1$ are the Schmidt coefficients. The maximum CHSH value attainable by this state is then
\begin{equation}
    S_{\mathrm{max}}
    =
    2\sqrt{1+4\lambda_0^2\lambda_1^2}.
    \label{eq:unrestricted_chsh}
\end{equation}

For an ideal GKP Bell state, the logical codewords are orthogonal and $\lambda_0=\lambda_1=1/\sqrt{2}$, recovering the Tsirelson bound $S_{\mathrm{max}}=2\sqrt{2}$. At finite squeezing, the nonzero overlap between the approximate GKP codewords modifies the Schmidt coefficients and reduces the maximum attainable violation. The difference between this unrestricted value and the homodyne-based benchmarks therefore quantifies the Bell nonlocality that remains inaccessible because of the imposed measurement constraints.

%The resulting measurement-constrained CHSH bounds are shown in Fig.~\ref{fig:figure1}(a). 
\ozlem{We compare the different measurement strategies, including homodyne-based and unrestricted strategies Fig.~\ref{fig:figure1}.} \ozlem{Figure~\ref{fig:figure1}(a) shows that} at finite squeezing, the accessible Bell violation depends strongly on the allowed measurement class. Periodic GKP binning provides the most restrictive \ozlem{CHSH value, with a violation emerging} only above approximately $2.6$~dB of GKP squeezing. Optimising the homodyne binning substantially improves the \ozlem{available} correlations, reducing the squeezing required for a violation to approximately $1$~dB and progressively approaching the unrestricted bound as squeezing increases. Furthermore, periodic GKP binning yields smaller CHSH violations than optimised homodyne binning up to approximately $8\,\mathrm{dB}$, beyond which the two strategies become nearly indistinguishable. This improvement arises from the finite-energy structure of the GKP codewords. At low squeezing, the broadened peaks and finite envelope shift the homodyne outcomes that contribute positively or negatively to the CHSH value away from the regular periodic GKP pattern as shown in the inset of Fig.~\ref{fig:figure1}(a). \ozlem{Optimising the binning adapts the classical outcome assignment to this structure, whereas at higher squeezing the optimal rule approaches periodic GKP binning (see Appendix~\ref{app:binning_strategies_figure}).} This highlights the importance of optimising the binning strategy at finite GKP squeezing \ozlem{as current optical demonstrations are limited to approximately $2.5~\mathrm{dB}$ of GKP squeezing~\cite{konno2024logical, larsen2025integrated}. Optimised binning can therefore provide a significant practical advantage for near-term optical demonstrations.}

In comparison, unrestricted measurements provide the maximum CHSH violation supported by the finite-energy state and rapidly approach the Tsirelson bound, $2\sqrt{2}$. It is important to note that the unrestricted measurements can yield a CHSH violation even at $0$~dB of GKP squeezing. This behaviour can be understood from the structure of the finite-energy GKP codewords in this limit.
\ozlem{At $s_{\mathrm{GKP}}=0\,\mathrm{dB}$, corresponding to $\Delta=1$, the Gaussian peaks $\ket{G_\Delta(q_0)}$ defined in Eq.~\eqref{eq:gkp_codewords} reduce to coherent states. Writing $\varepsilon=e^{-2\pi}$ and $\alpha=\sqrt{\pi/2}$, the finite-energy codewords can be expanded to first order in $\varepsilon$ as $\ket{\overline{0}_{\Delta=1}} \propto\ket{0}+\varepsilon\bigl(\ket{2\alpha}+\ket{-2\alpha}\bigr)$ and $\ket{\overline{1}_{\Delta=1}}\propto
\ket{\alpha}+\ket{-\alpha}$. After normalisation, the first-order correction to $\ket{\overline{0}_{\Delta=1}}$ does not contribute to the infidelity. The two codewords therefore have infidelities of order $\varepsilon^2\ll10^{-5}$ with the vacuum state $\ket{0}$ and the even cat state $\ket{C_+(\alpha)}$, respectively. We can therefore approximate}
\begin{equation}
    \ket{\bar{0}_{\Delta=1}}
    \simeq
    \ket{0},
    \;\;\;
    \ket{\bar{1}_{\Delta=1}}
    \simeq
    \ket{C_{+}(\alpha)},
    \;\;\;
    \alpha=\sqrt{\frac{\pi}{2}},
    \label{eq:gkp_0db_cat}
\end{equation}
where $\ket{C_+(\alpha)}$ is the even cat state defined later in Eq.~\eqref{eq:cat_codewords}. \ozlem{Numerically,} the overlap of these states with the corresponding finite-energy GKP codewords is greater than $0.99999$. Consequently, at $0$~dB the finite-energy GKP Bell state is effectively a vacuum-cat entangled state,
\begin{equation}
    \ket{\Phi_{\Delta=1}}
    \simeq
    \frac{1}{\sqrt{\mathcal N}}
    \left(
        \ket{0,0}
        +
        \ket{C_{+}(\alpha),C_{+}(\alpha)}
    \right).
    \label{eq:gkp_bell_0db}
\end{equation}
Despite the absence of GKP squeezing, this state remains entangled and can therefore violate the CHSH inequality under suitable unrestricted measurements. \ozlem{The resulting overlap between the codewords,~$c^2 =\vert \braket{0}{C_+(\alpha)} \vert^2= \sech(\alpha^2)$, leads to a concurrence~$\mathcal{C}_{\Delta=1}=(1-c^2)/(1+c^2) = \tanh^2(\pi/4)$ and a CHSH violation of~$S_{\Delta=1} = 2\sqrt{1+\mathcal{C}_{\Delta=1}^2}\approx 2.177$.}

To understand how the measurement constraints give rise to the different CHSH bounds in Fig.~\ref{fig:figure1}(a), we examine the corresponding effective logical measurements in Fig.~\ref{fig:figure1}(b). In the orthonormal logical basis, Bob's effective observable can be decomposed as
\begin{equation}
    B_0=c_I I+c_X X_L+c_Z Z_L,
    \label{eq:decomposition}
\end{equation}
where $X_L$ and $Z_L$ are the Pauli matrices acting on the two-dimensional logical subspace. The $Y_L$ component vanishes for the measurements considered here. For an ideal maximally entangled qubit pair, Bob's optimal CHSH measurement is $B_0=(X_L+Z_L)/\sqrt{2}$, corresponding to $c_X=c_Z=1/\sqrt{2}$ and $c_I=0$. For a finite-energy GKP Bell state, however, the optimal logical direction generally differs from this ideal tilted measurement, as reflected by the unrestricted coefficients in Fig.~\ref{fig:figure1}(b). As the GKP squeezing increases, the finite-energy codewords approach the ideal GKP limit and the unrestricted measurement converges towards $c_X=c_Z=1/\sqrt{2}$.

The homodyne-restricted measurements exhibit a further deviation from this finite-energy optimum. Periodic GKP binning approaches the unrestricted measurement only gradually as the squeezing increases, consistent with its reduced CHSH violation in Fig.~\ref{fig:figure1}(a). Optimising the homodyne binning substantially modifies the effective logical observable. At low squeezing, the optimum corresponds to the constant-output strategy $B_0\simeq I$, giving $c_I\simeq1$ and $c_X,c_Z\simeq0$ and explaining the $S=2$ plateau in Fig.~\ref{fig:figure1}(a). Above this regime, the optimum switches to a nontrivial logical measurement, with the identity contribution rapidly suppressed and the $X_L$ and $Z_L$ components approaching those of the unrestricted measurement.

\ozlem{Overall, these results show that optimising the homodyne binning is important in the finite-squeezing regime. It more than halves the GKP squeezing required to observe a CHSH violation, reducing the threshold from approximately $2.6\;\mathrm{dB}$ for periodic binning to approximately $1\;\mathrm{dB}$ for optimised binning. However, the low-squeezing limit also shows a fundamental limitation of homodyne-based measurements. Even though the finite-energy GKP Bell state supports a CHSH violation under unrestricted measurements, the optimised homodyne strategy remains at $S=2$. In this regime, the limiting factor for observing a CHSH violation is caused by the measurement constraint rather than the entanglement of the shared state. It is also important to note that the bounds in Fig.~\ref{fig:figure1} are logical-measurement benchmarks. The setting-dependent logical rotations used to define Bob's observables are generally non-Clifford, as shown in Fig.~\ref{fig:figure1}(b). In a fault-tolerant GKP architecture, such operations are commonly implemented by gate teleportation using an ancillary GKP-encoded magic state~\cite{baragiola2019all, yamasaki2020cost, konno2021non}. The finite quality of this additional resource is not included in these bounds. Accounting for its preparation and injection errors would further reduce the attainable CHSH values. This provides further motivation for the physical measurement strategies considered in the following section.}

\section{Revealing GKP Bell Nonlocality Without Logical Non-Clifford Gates}
\begin{figure*}[htbp]
%\hspace*{-0.3cm}
\includegraphics[scale=0.46]{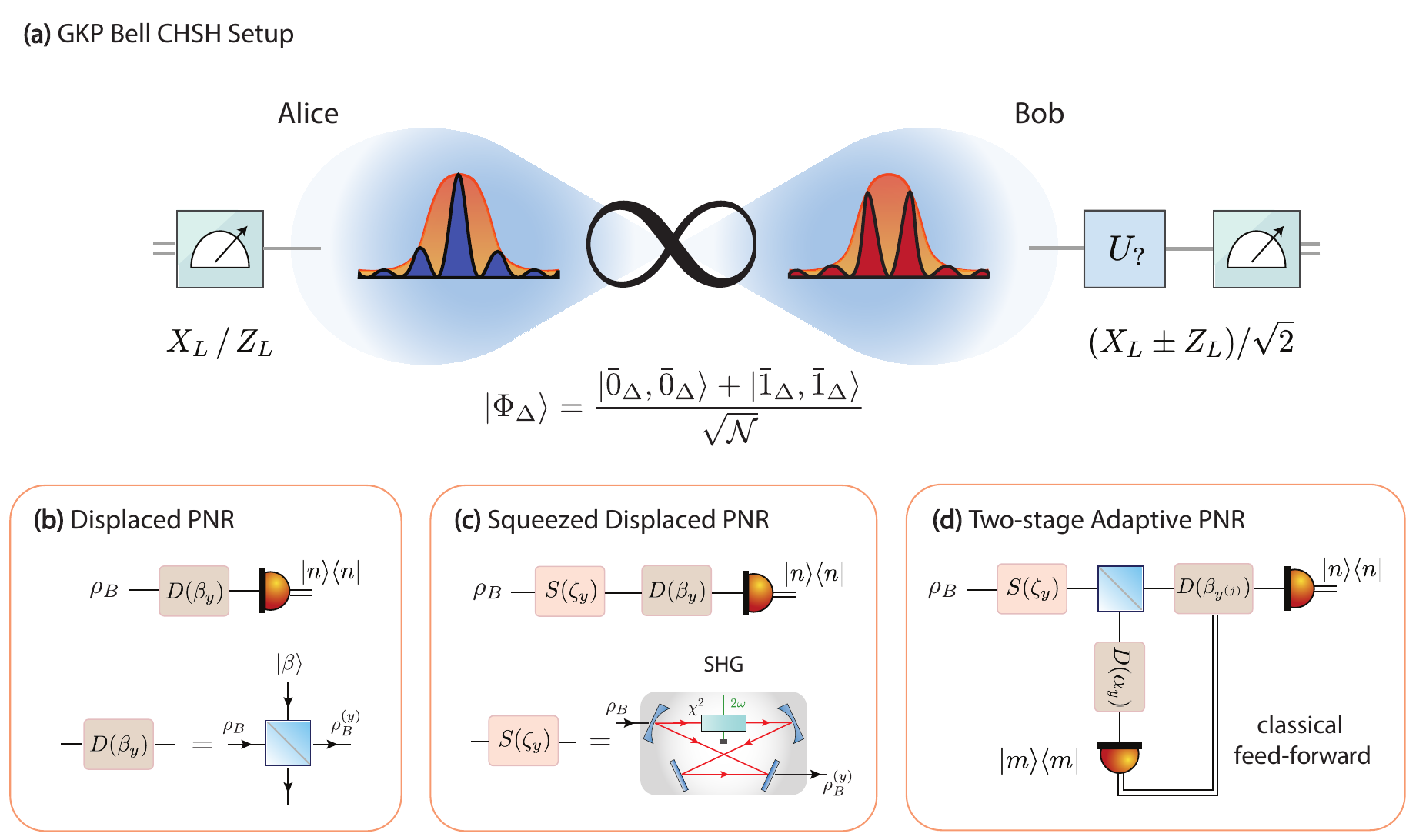}\hspace*{-0.1cm}
\caption{\label{fig:figure2}
GKP Bell-test setup and photon-number-resolving receiver architectures. (a) Alice and Bob share the logical GKP Bell state $\ket{\Phi_\Delta}$. Alice measures either $X_L$ or $Z_L$, while the optimal CHSH measurements for Bob correspond to the tilted logical observables $(X_L\pm Z_L)/\sqrt{2}$. The operation $U_{?}$ represents the local basis transformation required to access these observables. (b) In the displaced-PNR receiver, a setting-dependent displacement $D(\beta_y)$ is applied before photon-number detection. (c) The displaced-squeezed PNR receiver includes both a displacement $D(\beta_y)$ and a single-mode squeezing operation $S(\zeta_y)$ before detection. (d) In the two-stage adaptive receiver, the input is first squeezed and then divided at a beam splitter. The tapped mode is displaced by $D(\alpha_y)$ and measured, producing the outcome $m$. This result selects the conditional displacement $D(\beta_y^{(j)})$ applied to the retained mode before the second photon-number measurement, which produces $n$. For the two-branch receiver, $j=0$ when $m=0$ and $j=1$ when $m\geq1$. The complete photon-counting record $(m,n)$ is kept for the final optimised binary assignment.
}
\end{figure*}
The measurement-constrained bounds in Fig.~\ref{fig:figure1} demonstrate that substantially stronger Bell violations are available than those accessible through standard GKP measurements. We now investigate how these correlations can be reached using experimentally motivated measurements without implementing logical non-Clifford operations. \ozlem{Throughout the hybrid measurement strategies, Alice directly measures the logical $X_L$ and $Z_L$ observables using optimal binned homodyne detection, which requires no logical non-Clifford operations. As an alternative to implementing Bob’s tilted logical measurements using non-Clifford operations, we consider PNRD-based measurements combined with Gaussian preprocessing and adaptive detection, before extending these strategies to bilateral PNRD, as illustrated schematically in Fig.~\ref{fig:figure2}.}
%We consider measurement strategies based on homodyne and photon-number-resolving detection~(PNRD), together with Gaussian preprocessing and adaptive measurements, as illustrated schematically in Fig.~\ref{fig:figure2}.

\subsection{Displaced Photon-Number Detection}
We first consider displaced photon-number-resolving detection (PNRD) as a non-Gaussian measurement strategy for Bob. Throughout this section, Alice is restricted to homodyne detection with optimised binary binning, as introduced in the previous section. To increase the accessible measurement space, we precede PNRD by a phase-space displacement $D(\alpha)$, as illustrated in Fig.~\ref{fig:figure2}(b). The displacement can be expressed as $D(\alpha)=\exp\!\left(\alpha \hat{a}^{\dagger}-\alpha^{*}\hat{a}\right)$ where $\hat{a}=(\hat{q}+i\hat{p})/\sqrt{2}$ and $\hat{a}^{\dagger}=(\hat{q}-i\hat{p})/\sqrt{2}$ are the annihilation and creation operators, respectively with $[\hat{q},\hat{p}]=i$. For Bob's measurement setting $y$, the displacement $\beta_y$ transforms his state according to
\begin{equation}
    \rho_B
    \longrightarrow
    \rho_B^{(y)}
    =
    D(\beta_y)\rho_BD^{\dagger}(\beta_y),
    \label{eq:displaced_state}
\end{equation}
after which PNRD is performed in the Fock basis $\{\ket{n}\}$. The probability of detecting $n$ photons is therefore
\begin{equation}
    p(n|y)
    =
    \bra{n}
    D(\beta_y)\rho_BD^{\dagger}(\beta_y)
    \ket{n}.
    \label{eq:displaced_pnr_probability}
\end{equation}
To obtain the binary outcomes required for the CHSH test, each photon-number outcome $n$ is assigned a value $s_{y,n}\in\{-1,+1\}$. The corresponding dichotomic observable, expressed on Bob's state before the displacement, is
\begin{equation}
    B_y^{\mathrm{PNR}}
    (\beta_y,\mathbf{s}_y)
    =
    D^{\dagger}(\beta_y)
    \left(
        \sum_{n=0}^{N_{\mathrm{cut}}-1}
        s_{y,n}\ket{n}\!\bra{n}
    \right)
    D(\beta_y),
    \label{eq:displaced_pnr}
\end{equation}
where $\mathbf{s}_y=\{s_{y,n}\}$ denotes the photon-number binning for measurement setting $y$. The cutoff $N_{\mathrm{cut}}$ is chosen sufficiently large that the probability contained in the neglected higher-photon-number subspace is negligible for all parameters considered.

For a fixed displacement, the contribution of each photon-number outcome to the CHSH parameter is determined by
\begin{equation}
    g_{y,n}(\beta_y)
    =
    \bra{n}
    D(\beta_y) C_y D^{\dagger}(\beta_y)
    \ket{n},
    \label{eq:pnr_score}
\end{equation}
where $C_y$ is Bob's CHSH correlation operator. Since the binary assignment of each photon-number outcome can be chosen independently, the optimal binning is obtained directly as
\begin{equation}
    s_{y,n}^{\mathrm{opt}}
    =
    \operatorname{sign}
    \left[g_{y,n}(\beta_y)\right].
    \label{eq:pnr_optimal_binning}
\end{equation}
The photon-number binning can therefore be optimised analytically, leaving only the complex displacement $\beta_y$ to be optimised numerically.

\subsection{Gaussian-Preprocessed PNR Detection}
We next enlarge the accessible measurement space by allowing single-mode Gaussian preprocessing prior to PNRD. We first keep optimised homodyne binning for Alice and apply the Gaussian-preprocessed PNR measurement only to Bob. We subsequently consider a symmetric strategy in which both Alice and Bob employ the same measurement architecture, with their respective measurement parameters optimised independently. 

In addition to the phase-space displacement considered above, Bob applies a single-mode squeezing operation before photon-number detection as shown in Fig.~\ref{fig:figure2}(c). Squeezing has previously been shown to enhance Bell violations of entangled superpositions of coherent states when combined with photon-counting measurements, with the improvement depending on the measurement scheme and squeezing direction~\cite{lee2009effects}. Motivated by these results, we supplement the phase-space displacement considered above with a single-mode squeezing operation and investigate whether this additional Gaussian degree of freedom can similarly enhance the Bell violation of finite-energy GKP states. For measurement setting $y$, the Gaussian preprocessing is described by $U_y^{\mathrm{G}}=D(\beta_y)S(\zeta_y)$, where $\beta_y\in\mathbb{C}$ is the displacement amplitude and $S(\zeta_y)=\exp\!\left[ \frac{1}{2}\left(\zeta_y^{*}\hat{a}^{2}-\zeta_y\hat{a}^{\dagger 2}\right)\right]$ is the single-mode squeezing operator, 
%with $\zeta_y=r_y e^{i\phi_y}$. 
with $\zeta_y \in \mathbb{C}$ the complex squeezing parameter.
\ozlem{In the numerical optimisation, we restrict the squeezing parameter to a fixed quadrature axis, taking $\zeta_y\in[-1,1]\subset\mathbb{R}$. Positive and negative values select orthogonal squeezing axes. The reported values are therefore the best-found CHSH values within this real-squeezing family.} The physical sequence therefore consists of squeezing Bob's mode, applying the displacement, and subsequently performing PNRD. The state immediately before detection is
\begin{equation}
    \rho_B^{(y)}
    =
    D(\beta_y)S(\zeta_y)
    \rho_B
    S^{\dagger}(\zeta_y)D^{\dagger}(\beta_y).
    \label{eq:gaussian_preprocessed_state}
\end{equation}
As in the displaced-PNRD strategy, each photon-number outcome is assigned a binary value $s_{y,n}\in\{-1,+1\}$. The corresponding dichotomic observable, expressed on the state prior to the Gaussian preprocessing, is
\begin{equation}
    B_y^{\mathrm{G\text{-}PNR}}
    \!=\!
    S^{\dagger}(\zeta_y)D^{\dagger}(\beta_y)
    \left(\!
        \sum_{n=0}^{N_{\mathrm{cut}}-1}
        s_{y,n}\ket{n}\!\bra{n}
    \!\right)
    D(\beta_y)S(\zeta_y).
    \label{eq:gaussian_pnr_observable}
\end{equation}
For each measurement setting $y$, a single set of Gaussian parameters $(\beta_y,\zeta_y)$ is optimised to maximise the CHSH parameter and is fixed for all photon-number outcomes. For fixed $(\beta_y,\zeta_y)$, the optimal binary assignment $s_{y,n}$ is then obtained independently for each photon-number outcome from the sign of its contribution to the CHSH parameter. 
%The binary assignment can therefore be optimised analytically, while the displacement and squeezing parameters are optimised numerically.
\ozlem{The observable in Eq.~\eqref{eq:gaussian_pnr_observable} is an analytical solution of the optimal binary assignment, whereas the displacement and squeezing parameters are optimised numerically.}

\subsection{Adaptive PNR Measurements}
\label{sec:adaptive_pnr}
We finally consider an adaptive measurement strategy, in which the outcome of an initial photon-number measurement is used to condition the subsequent measurement performed on Bob's mode. Adaptive photon-counting receivers have previously been explored in optical state discrimination, where measurement outcomes obtained at an earlier stage are used to determine the displacement
applied at a subsequent stage~\cite{becerra2013implementation,cui2022quantum}. We first \ozlem{fix} the optimised homodyne binning for Alice and apply the adaptive receiver to Bob alone. We then consider the symmetric case \ozlem{where} both parties employ the adaptive PNR measurement, with independently optimised receiver parameters.

For each measurement setting $y$, Bob first applies a single-mode squeezing operation $S(\zeta_y)$ to his state and then taps off a fraction of the optical mode by mixing it with vacuum on a beam splitter as illustrated in Fig.~\ref{fig:figure2}(d). The resulting two-mode state is
\begin{equation}
    \rho_{Bv}^{(y)}
    =
    U_{\mathrm{BS}}(T_y)
    \left[
        S(\zeta_y)\rho_B S^\dagger(\zeta_y)
        \otimes
        \ket{0}\!\bra{0}_{v}
    \right]
    U_{\mathrm{BS}}^\dagger(T_y),
    \label{eq:adaptive_bs_state}
\end{equation}
where $T_y\in[0,1]$ is the beam-splitter transmissivity for measurement setting $y$, and $\hat v$ denotes the annihilation operator of the ancillary vacuum mode. We use the convention $U_{\mathrm{BS}}(T_y) = \exp\!\left[\cos^{-1}\big(\sqrt{T_y}\big)\left(\hat{a}^{\dagger}\hat{v}-\hat{a}\hat{v}^{\dagger}\right)\right]$. One output mode is kept by Bob, while the other constitutes the tapped mode used for the first stage of the adaptive measurement. The tapped mode is displaced by $D(\alpha_y)$ and subsequently measured using PNRD, yielding the photon-number outcome $m$. Immediately prior to this first detection, the joint state is therefore
\begin{equation}
    \tilde{\rho}_{Bv}^{(y)}
    =
    \left[
        I\otimes D(\alpha_y)
    \right]
    \rho_{Bv}^{(y)}
    \left[
        I\otimes D^\dagger(\alpha_y)
    \right].
    \label{eq:adaptive_first_displacement}
\end{equation}
Detection of $m$ photons in the tapped mode projects the retained mode onto the conditional state
\begin{equation}
    \rho_{B|m,y}
    =
    \frac{
        \operatorname{Tr}_{v}
        \left[
            \left(
                I\otimes\ket{m}\!\bra{m}
            \right)
            \tilde{\rho}_{Bv}^{(y)}
        \right]
    }{
        p(m|y)
    },
    \label{eq:adaptive_conditional_state}
\end{equation}
where
\begin{equation}
    p(m|y)
    =
    \operatorname{Tr}
    \left[
        \left(
            I\otimes\ket{m}\!\bra{m}
        \right)
        \tilde{\rho}_{Bv}^{(y)}
    \right]
\end{equation}
is the probability of obtaining the first-stage outcome $m$. The first-stage outcome is then used to condition the displacement applied to the kept mode. We employ a two-branch feed-forward strategy,
\begin{equation}
    \beta_y(m)
    =
    \begin{cases}
        \beta_y^{(0)}, & m=0,\\[4pt]
        \beta_y^{(1)}, & m>0.
    \end{cases}
    \label{eq:adaptive_displacement}
\end{equation}
\ozlem{This choice distinguishes a no-click event from the detection of one or more photons in the tapped mode, while avoiding a separate feed-forward displacement for every resolved photon number. Then} the kept state is transformed according to
\begin{equation}
    \rho_{B|m,y}
    \longrightarrow
    D\!\left[\beta_y(m)\right]
    \rho_{B|m,y}
    D^\dagger\!\left[\beta_y(m)\right].
    \label{eq:adaptive_second_displacement}
\end{equation}
A second PNR measurement is subsequently performed on the kept mode, yielding the photon-number outcome $n$.

The complete measurement record for Bob is therefore given by the pair $(m,n)$. To obtain the binary outcome required for the CHSH test, each detection record is assigned a value $s_{y,mn}\in\{-1,+1\}$.
\begin{figure*}[htbp]
%\hspace*{-0.3cm}
\includegraphics[scale=0.42]{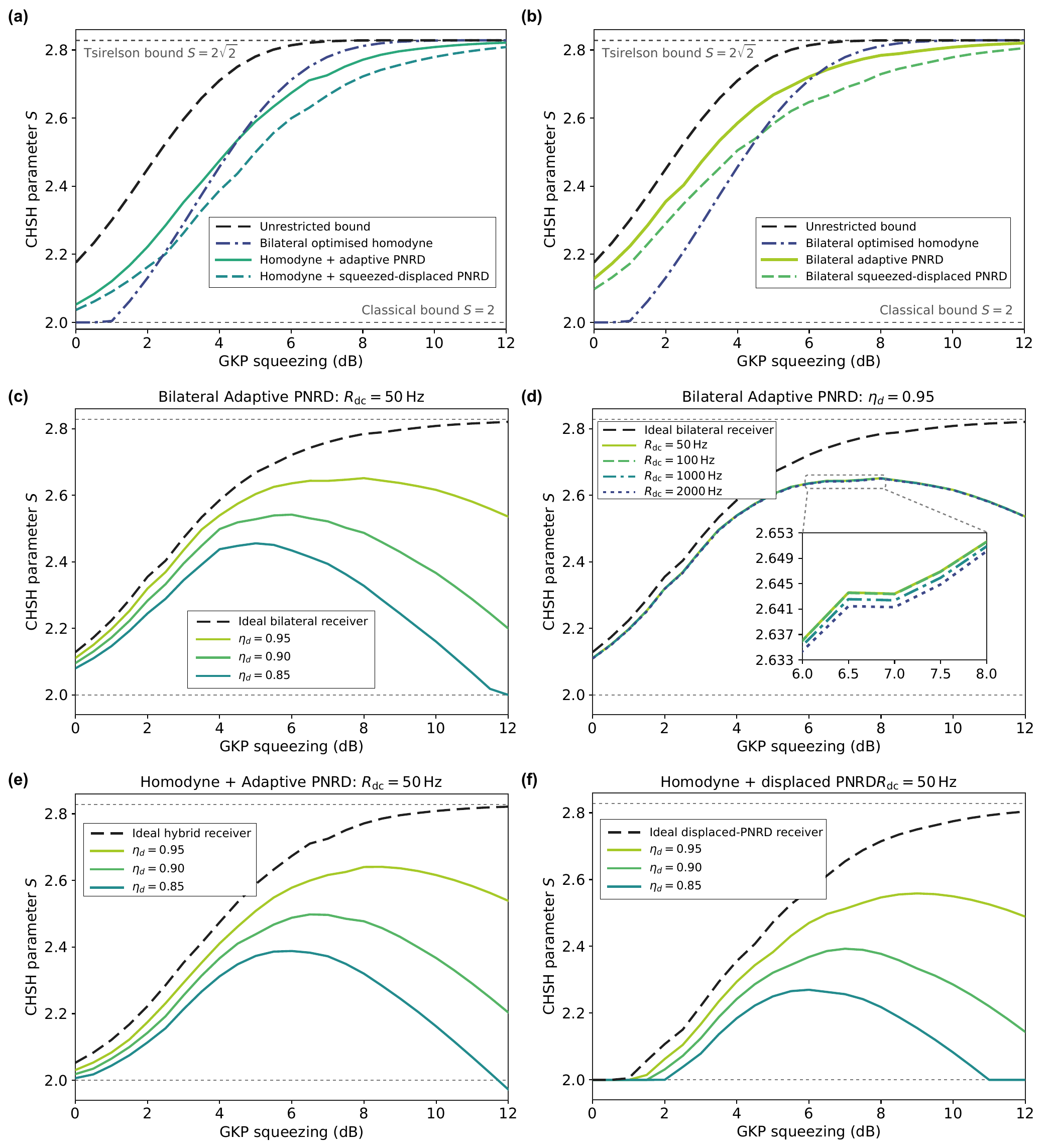}\hspace*{-0.24cm}
\caption{\label{fig:figure3}
Physical measurement strategies and detector-noise robustness for finite-energy GKP Bell states. (a) The unrestricted dichotomic benchmark, bilateral optimised homodyne detection, and homodyne detection for Alice combined with either adaptive PNRD or squeezed-displaced PNRD for Bob. (b) The unrestricted benchmark, bilateral optimised homodyne detection, bilateral adaptive PNRD, and bilateral squeezed-displaced PNRD without a tapped mode. In the adaptive receiver, the first PNR outcome selects the second displacement according to $m=0$ or $m\geq1$. Panels (c)--(f) include detector imperfections. (c) Bilateral adaptive PNRD for $\eta_d=0.95$, $0.90$, and $0.85$ at $R_{\mathrm{dc}}=50\,\mathrm{Hz}$. (d) Dark-count dependence at $\eta_d=0.95$ for the bilateral adaptive PNRD scheme. The inset enlarges the $6$--$8\,\mathrm{dB}$ region. (e) Homodyne plus adaptive PNRD and (f) homodyne plus displaced PNRD, each for the same PNRD efficiencies. The dashed black curves in (c), (e), and (f) show the corresponding ideal physical receivers. Noisy PNRD calculations use a $100\,\mathrm{ns}$ gate; in (e) and (f), $\eta_{\mathrm{hom}}=0.995$ with $20\,\mathrm{dB}$ electronic-noise clearance. Horizontal lines indicate $S=2$ and $S=2\sqrt{2}$.
}
\end{figure*}

For fixed optical parameters, the optimal assignment is determined by the contribution of each detection record to the CHSH parameter. Denoting the POVM element associated with the outcome $(m,n)$ by $E_{y,mn}$, we define
\begin{equation}
    d_y(m,n)
    =
    \operatorname{Tr}
    \left[
        C_y E_{y,mn}
    \right],
    \label{eq:adaptive_score}
\end{equation}
where $C_y$ is Bob's corresponding CHSH correlation operator. Since each detection record can be assigned independently to either binary outcome, the optimal assignment is
\begin{equation}
    s_{y,mn}^{\mathrm{opt}}
    =
    \operatorname{sign}
    \left[
        d_y(m,n)
    \right].
    \label{eq:adaptive_optimal_assignment}
\end{equation}
The resulting dichotomic observable is therefore
\begin{equation}
    B_y^{\mathrm{ad}}
    =
    \sum_{m,n}
    s_{y,mn}^{\mathrm{opt}} E_{y,mn},
    \label{eq:adaptive_observable}
\end{equation}
where $E_{y,mn}$ is the POVM element associated with detecting $m$ photons at the first PNR detector and $n$ photons at the second PNR detector for measurement setting $y$. 
%The set $\{E_{y,mn}\}$ describes the complete adaptive measurement on Bob's input mode and satisfies
%\begin{equation}
%    p(m,n|y)=\operatorname{Tr}\!\left[\rho_B E_{y,mn}\right],\qquad\sum_{m,n}E_{y,mn}=I.
%\end{equation}

For each measurement setting $y$, the receiver parameters $\{\zeta_y,T_y,\alpha_y,\beta_y^{(0)},\beta_y^{(1)}\}$ are jointly optimised to maximise the CHSH parameter and remain fixed throughout the measurement. The dependence on the first-stage outcome $m$ enters only through the pre-optimised feed-forward rule in Eq.~\eqref{eq:adaptive_displacement}. For fixed receiver parameters, the final binary assignment $s_{y,mn}^{\mathrm{opt}}$ is determined independently for each detection record $(m,n)$.

Figure~\ref{fig:figure3}(a) compares the CHSH values obtained with different photon-counting receiver architectures on Bob’s side, with Alice’s measurement fixed to homodyne detection. The simplest hybrid strategy, in which Alice performs optimised homodyne detection and Bob performs displaced PNRD, remains below the bilateral optimised-homodyne curve over most of the squeezing range (\ozlem{see Appendix~\ref{app:parameters_used} for the numerical parameters used in these calculations and the corresponding optimised displacements shown in Fig.~\ref{fig:figure2_appendix}}). Introducing squeezing before Bob's displacement improves the violation at low GKP squeezing, but this advantage diminishes as the squeezing increases, with the receiver eventually falling below the optimised-homodyne curve (\ozlem{see Fig.~\ref{fig:figure3_appendix} for the corresponding optimised displacement and squeezing parameters}). A considerably larger improvement is obtained with the two-stage adaptive receiver, showing that conditioning the second displacement on the first photon-counting outcome provides a useful measurement resource beyond fixed Gaussian preprocessing.

\ozlem{In comparison, Fig.~\ref{fig:figure3}(b) considers the symmetric setting in which Alice and Bob employ the same measurement scheme.} The strongest measurement-constrained violation at low and intermediate squeezing is obtained when the adaptive receiver is employed bilaterally. Up to approximately $6.3\,\mathrm{dB}$, this strategy outperforms bilateral optimised homodyne detection and lies between the homodyne and unrestricted curves. Over the interval from $0$ to $6\,\mathrm{dB}$, the bilateral adaptive receiver achieves approximately $95\%$--$98\%$ of the unrestricted CHSH value, with the difference between the two curves remaining below $0.125$. At low squeezing, the bilateral squeezed-displaced PNRD receiver without a tapped mode gives the next-highest CHSH value among the measurement-constrained strategies.

The change in the ordering of the measurement strategies reflects the finite-energy nature of the GKP states. At low GKP squeezing, the broadened GKP peaks make it difficult for homodyne binning to access the logical correlations required for a strong CHSH violation. The PNRD receivers \ozlem{can instead produce} larger violations through Gaussian preprocessing and adaptive feed-forward. As the GKP squeezing increases, the state approaches the ideal GKP code space and optimised homodyne binning increasingly approximates the required logical measurements. Bilateral optimised homodyne detection consequently approaches the unrestricted value. Achieving this high-squeezing benchmark, however, requires access to the tilted logical measurements associated with non-Clifford operations. The PNRD receivers provide a more experimentally achievable alternative: at finite GKP squeezing, they give CHSH values closer to the unrestricted result using Gaussian preprocessing, photon counting and simple feed-forward. Their strong performance at low squeezing shows that sizeable Bell violations can be obtained without highly squeezed GKP Bell states or logical non-Clifford operations.

\ozlem{We next show the performance of these measurement schemes in the presence of experimental imperfections. Figures~\ref{fig:figure3}(c)--(f) include PNRD inefficiency and dark counts, as well as homodyne inefficiency and electronic noise. We consider bilateral adaptive PNRD detection, Alice performing homodyne detection while Bob uses an adaptive PNRD, and Alice performing homodyne detection while Bob uses a displaced PNRD, since the latter is the least experimentally challenging scheme to implement. For the efficiency analysis, we use PNRD efficiencies of $95\%$, $90\%$, and $85\%$, with a fixed dark-count rate of $50\,\mathrm{Hz}$ (refer to Appendix~\ref{app:noisy_pnrd} for the method). The homodyne detectors are modelled with an efficiency of $99.5\%$ and an electronic-noise clearance of $20\,\mathrm{dB}$ (see Appendix~\ref{app:homodyne_efficiency} for the method), parameters that are within current experimental capabilities~\cite{gehring2015implementation, larsen2024continuous}. Similarly, photon-number-resolving detectors with efficiencies around $95\%$ and low dark-count rates have been demonstrated experimentally~\cite{li2024high, lita2008counting}. All measurement strategies yield lower CHSH values as the PNRD efficiency is reduced. They also exhibit an optimal GKP squeezing, around $6$--$8\,\mathrm{dB}$, beyond which the CHSH value deteriorates, as shown in Fig.~\ref{fig:figure3}(c), (e), and (f). Although all three strategies maintain CHSH violations over most of the efficiency range, their performance is more limited at $85\%$ efficiency. The bilateral adaptive strategy stops violating the CHSH inequality at $12\,\mathrm{dB}$, while the homodyne-adaptive PNRD strategy falls below the classical bound at approximately $11.6\,\mathrm{dB}$. In contrast, the homodyne-displaced PNRD strategy can only violate the inequality between $2$ and $11\;\mathrm{dB}$. This behaviour arises because increasing the GKP squeezing also increases the mean photon number of the state. At sufficiently high squeezing, a fixed PNRD inefficiency leads to a greater loss of photon-number information, which reduces the ability of the receivers to resolve the logical correlations required for a CHSH violation. The resulting loss-induced degradation eventually outweighs the improvement obtained from better approximating the ideal GKP code space. Furthermore, Fig.~\ref{fig:figure3}(d) shows that the CHSH violation of the bilateral adaptive PNRD receiver is largely insensitive to the dark-count rates considered in the panel. This is because the probability of a dark count within the $100\,\mathrm{ns}$ detection window is small, even at the largest rate considered. Detector inefficiency is more detrimental because photon-loss directly affects the incoming GKP states and degrades the photon-number information required to resolve their logical correlations.}

\section{Generalisation to entangled coherent states and cat-Bell states}
Having found that bilateral adaptive PNRD gives the strongest CHSH violation among the experimentally motivated measurement strategies considered for the finite-energy GKP Bell state, %we next ask whether its performance is specific to the GKP lattice structure or extends to coherent-state superpositions. 
\ozlem{one may ask whether this performance is specific to the GKP lattice structure or extends to coherent-state superpositions. In this section, we show that this strategy also gives strong CHSH violations for coherent-state superpositions.} We apply the same two-branch bilateral adaptive receiver to two distinct resources: a cat-code Bell state constructed from the orthogonal even- and odd-parity cat states, and an entangled coherent state constructed directly from the nonorthogonal coherent states \(\ket{\alpha}\) and \(\ket{-\alpha}\). Our aim is to compare the Bell nonlocality accessible from different bosonic states when both the local measurement architecture and the mean energy per mode are held fixed.

We first consider a Bell pair encoded in the cat-state basis. Without loss of generality, we take $\alpha>0$ to be real and define the normalised even and odd cat states as
\begin{equation}
    \ket{C_{\pm}(\alpha)}
    =
    \frac{\ket{\alpha}\pm\ket{-\alpha}}
    {\sqrt{2\left(1\pm e^{-2\alpha^2}\right)}}.
    \label{eq:cat_codewords}
\end{equation}
Although the coherent states $\ket{\alpha}$ and $\ket{-\alpha}$ are not orthogonal at finite $\alpha$, their even and odd superpositions occupy disjoint photon-number-parity sectors and therefore form an exactly orthogonal basis. We use this basis to construct the cat-code Bell state
\begin{equation}
    \ket{\Phi_{\mathrm{cat}}(\alpha)}
    =
    \frac{
        \ket{C_+(\alpha),C_+(\alpha)}
        +
        \ket{C_-(\alpha),C_-(\alpha)}}{\sqrt{2}}.
    \label{eq:cat_bell_pair}
\end{equation}
The two terms have equal Schmidt weights, so this state is maximally entangled
\AD{($\mathcal{C}=1$)}
for every \(\alpha>0\)
and its unrestricted CHSH value is the Tsirelson bound, $S_{\max}=2\sqrt{2}$.
\AD{For instance, in the
low amplitude regime,
the cat code Bell state~$\ket{\Phi_\mathrm{cat}}$ approaches~$\nicefrac{1}{\sqrt{2}}(\ket{0,0} + \ket{1,1})$
as~$\alpha\to0$.}

We next consider the even ECS, obtained by coherently superposing two correlated coherent-state product states,
\begin{equation}
    \ket{\Psi_{\mathrm{ECS}}(\alpha)}
    =
    \frac{
        \ket{\alpha,\alpha}+\ket{-\alpha,-\alpha}
    }{
        \sqrt{2\left(1+e^{-4\alpha^2}\right)}
    }.
    \label{eq:ecs_state}
\end{equation}
The normalization explicitly accounts for the nonzero two-mode overlap
$\langle{\alpha,\alpha|{-\alpha},{-\alpha}}\rangle=e^{-4\alpha^2}$. Consequently, the two components of the ECS do not define orthogonal local states at finite $\alpha$, and the state is not maximally entangled. As $\alpha$ increases, the overlap is exponentially suppressed and the ECS becomes asymptotically equivalent to the maximally entangled cat-code Bell state in Eq.~\eqref{eq:cat_bell_pair}.

The origin of this reduced entanglement becomes transparent when the ECS is expressed in the orthogonal cat basis. Setting $s=e^{-2\alpha^2}$, its Schmidt decomposition is
\begin{equation}
    \ket{\Psi_{\mathrm{ECS}}(\alpha)}
    =
    \lambda_+\ket{C_+(\alpha),C_+(\alpha)}
    +
    \lambda_-\ket{C_-(\alpha),C_-(\alpha)},
    \label{eq:ecs_schmidt_form}
\end{equation}
where $\lambda_{\pm}=\frac{1\pm s}{\sqrt{2(1+s^2)}}$. The corresponding concurrence and state-dependent CHSH ceiling are
\begin{align}
    &\mathcal{C}_{\mathrm{ECS}}
    =2\lambda_+\lambda_-
    =\frac{1-s^2}{1+s^2}
    =\tanh(2\alpha^2),
    \nonumber \\
    &S_{\mathrm{ECS}}^{\max}
    =2\sqrt{1+\mathcal{C}_{\mathrm{ECS}}^2}.
    \label{eq:ecs_chsh_ceiling}
\end{align}
Unlike the cat-code Bell pair, this ceiling lies below $2\sqrt{2}$ at finite $\alpha$ and approaches the Tsirelson bound only in the large-amplitude limit.
\AD{In the low amplitude limit~($\alpha \ll 1$),
the ECS becomes~$\ket{\Psi_\mathrm{ECS}}\propto \ket{0,0} + \alpha^2 \left ( \ket{1,1} 
 + \frac{1}{\sqrt{2}} \left ( \ket{2,0} + \ket{0,2} \right )
\right ) $,
which approaches a separable state as~$\alpha \to 0$. By contrast, replacing the plus
sign in Eq.~\eqref{eq:ecs_state} with a minus sign gives an odd ECS that approaches $(|1,0\rangle+|0,1\rangle)/\sqrt{2}$ as $\alpha\to0$. The odd ECS is maximally entangled for every $\alpha>0$ and therefore has unrestricted CHSH value $2\sqrt{2}$. It is locally equivalent to the cat-code Bell pair considered above. We use the even ECS to examine the distinct effect of
amplitude-dependent entanglement on the Bell violation.}

\subsection{Measurement strategies}
\begin{figure*}[htbp]
%\hspace*{-0.3cm}
\includegraphics[scale=0.48]{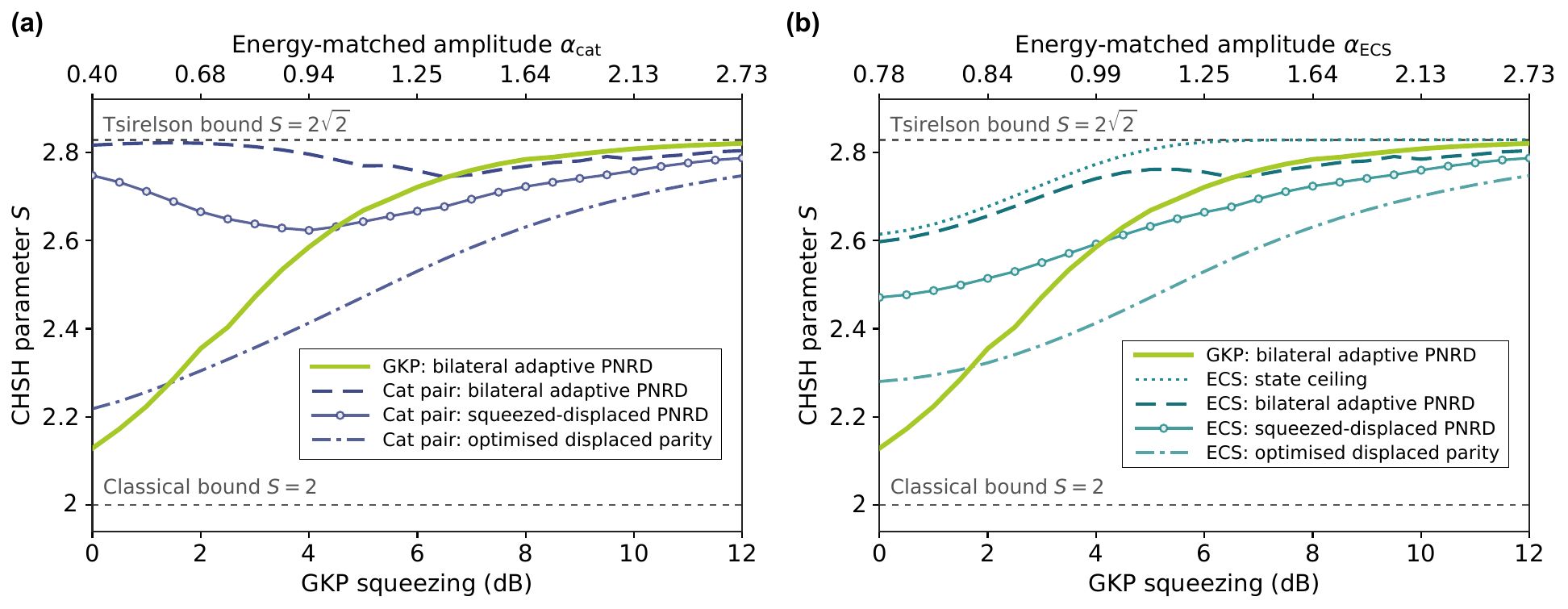}\hspace*{-0.1cm}
\caption{\label{fig:figure4}
Energy-matched comparison of the finite-energy GKP Bell state with coherent-state Bell resources. Panel (a) shows the cat-code Bell pair and panel (b) the entangled coherent state (ECS), each compared with the bilateral adaptive-PNRD result for the GKP Bell state. At each value of GKP squeezing, the mean photon number per mode is matched across the three state families, with the corresponding cat-state and ECS amplitudes shown on the upper axes. The yellow-green curve shows the bilateral adaptive-PNRD result for the GKP Bell state. Blue curves in (a) and teal curves in (b) correspond to the cat-code Bell pair and ECS, respectively. For each coherent-state family, the dashed curve shows the two-branch adaptive-PNRD receiver, the open-circle curve shows squeezed-displaced PNRD without a tapped mode, and the dash-dotted curve shows displaced parity with displacements optimised at each amplitude. The dotted teal curve in (b) gives the state-dependent ECS CHSH ceiling. The horizontal lines indicate the classical bound $S=2$ and the Tsirelson bound $S=2\sqrt{2}$, which is also the constant ceiling for the cat-code Bell pair.
}
\end{figure*}
To compare the cat-code Bell state and the ECS under identical measurement constraints, we apply the bilateral adaptive PNR receiver introduced in Sec.~\ref{sec:adaptive_pnr} to both states. Alice and Bob implement the observables $A_x^{\mathrm{ad}}$ and $B_y^{\mathrm{ad}}$, respectively, where $x$ and $y$ label their CHSH settings. %The receiver parameters are optimised independently for each party and setting. 
\ozlem{The receiver parameters are distinct for each party and setting and are optimised by alternating Alice and Bob updates using the see-saw procedure described in Appendix~\ref{app:measurement_optimisation}.} For Bob, these are $\zeta_y$, $T_y$, $\alpha_y$, $\beta_y^{(0)}$, and $\beta_y^{(1)}$, as defined in the previous section; Alice uses the corresponding $x$-labelled parameters.

The feed-forward is restricted to two branches: the outcome $m=0$ selects $\beta_y^{(0)}$, while any outcome $m>0$ selects $\beta_y^{(1)}$. 
%This grouping affects only the choice of conditional displacement. The first detector remains photon-number resolving, and the complete record $(m,n)$ is retained.
\ozlem{The complete PNR record $(m,n)$ is kept for the final binary outcome assignment.} Each record is then assigned a binary value according to the optimal sign rule in Eq.~\eqref{eq:adaptive_optimal_assignment}, with assignments $s_{x,mn}^{\mathrm{opt}}$ for Alice and $s_{y,mn}^{\mathrm{opt}}$ for Bob.

Following established Bell tests of coherent-state superpositions and ECSs based on phase-space displacement followed by photon-number-parity detection~\cite{jeong2003quantum,lee2009effects}, we take displaced parity as our second measurement
\begin{equation}
    \Pi(\gamma)
    =
    D^{\dagger}(\gamma)
    \left(
        \sum_{n=0}^{\infty}(-1)^n\ket{n}\!\bra{n}
    \right)
    D(\gamma).
    \label{eq:displaced_parity_observable}
\end{equation}
For the symmetric states considered here, the displacements can be restricted to the imaginary axis. We therefore choose
\begin{align}
    &A_0^{\Pi}=\Pi(0),
    \qquad
    A_1^{\Pi}=\Pi(i\gamma_A),
    \nonumber \\
    &B_0^{\Pi}=\Pi(-i\gamma_B),
    \;
    B_1^{\Pi}=\Pi(i\gamma_B),
    \label{eq:cat_displaced_parity_settings}
\end{align}
where $\gamma_A$ and $\gamma_B$ are optimised independently at each amplitude, rather than fixed to their large-$\alpha$ asymptotic values.

Finally, we consider bilateral squeezed-displaced PNRD without a tapped mode. This is the bilateral version of the Gaussian-preprocessed observable $B_y^{\mathrm{G\text{-}PNR}}$ in Eq.~\eqref{eq:gaussian_pnr_observable}, together with the corresponding observable $A_x^{\mathrm{G\text{-}PNR}}$ for Alice. The beamsplitter and first PNR detection are omitted, so each party applies only $S(\zeta)$ followed by $D(\beta)$ and PNRD. The displacement, squeezing, and photon-number signs are optimised independently for every setting.

Each of these strategies defines a valid and experimentally \ozlem{feasible} CHSH test. \ozlem{The measurement families form a hierarchy,
$
\text{displaced parity}
\subset
\text{displaced PNRD}
\subset
\text{squeezed-displaced PNRD}
\subset
\text{adaptive PNRD}.
$ Displaced parity is recovered by a fixed photon-number-parity assignment, while squeezed-displaced PNRD is recovered from the adaptive receiver by removing the tapped mode and making the two conditional displacements identical. Thus, under a global optimisation, $S_{\mathrm{ad}}\geq S_{\mathrm{G\text{-}PNR}}\geq S_{\Pi}$. Here, we report the largest CHSH value found numerically for each receiver
at each energy, without a guarantee of global optimality.} %Therefore,} we report the largest achieved value at each energy,
%\begin{equation}
%    S_{\mathrm{best}}
%    =
%    \max\!\left\{
%        S_{\mathrm{ad}},
%        S_{\Pi},
%        S_{\mathrm{G\text{-}PNR}}
%    \right\}.
%    \label{eq:cat_best_measurement_envelope}
%\end{equation}
%This envelope does not combine different receivers within a single Bell trial.
%Rather, for a fixed state amplitude, the receiver giving the largest certified lower bound is selected before the experiment is performed.
%\ozlem{Instead, it is meant to identify the optimal choice of receiver, given a state amplitude,  before an experiment is performed.}

\subsection{Bell-inequality violation}
Figure~\ref{fig:figure4} compares the three physical measurement strategies with the state-dependent CHSH \ozlem{values} \ozlem{(see Appendix~\ref{app:energy_matching} for the energy matching of the states)}. For the cat-code Bell pair, the unrestricted ceiling remains $2\sqrt{2}$ over the full energy range. For the ECS, the ceiling follows Eq.~\eqref{eq:ecs_chsh_ceiling} and converges to $2\sqrt{2}$ as the coherent components become orthogonal. 
%The individual receiver curves identify which physical measurement is responsible for the best achieved value in Eq.~\eqref{eq:cat_best_measurement_envelope}.

The cat-code Bell pair gives a substantially larger violation than the bilateral adaptive-PNRD strategy for the energy-matched GKP Bell state at low and intermediate energies \ozlem{as shown in Fig.~\ref{fig:figure4}(a)}. This is expected because the cat codewords have opposite photon-number parity, making photon-number-resolving detection naturally suited to distinguishing their logical structure. \ozlem{By comparison, low-squeezing GKP states have broadened, overlapping lattice peaks whose logical correlations are less directly resolved by this receiver.} The ECS is also well matched to photon counting, although its finite-amplitude CHSH ceiling \ozlem{is lower because the nonzero overlap between $\ket{\alpha}$ and $\ket{-\alpha}$ prevents it from being maximally entangled (see Fig.~\ref{fig:figure4}(b)).}

\ozlem{The measurement hierarchy is clearly visible for both coherent-state families as shown in Fig.~\ref{fig:figure4}. Bilateral adaptive PNRD gives the largest CHSH
value across the plotted amplitude range, followed by squeezed-displaced PNRD and then optimised displaced parity. The adaptive and squeezed-displaced receivers exhibit a non-monotonic amplitude dependence, reflecting changes in the overlap and interference of the coherent-state components. Transitions between locally optimal solutions of the nonconvex receiver optimisation may also contribute to this behaviour. By contrast, the optimised displaced-parity value increases more steadily as the coherent-state components become more distinguishable in phase space. Furthermore, both the cat-code Bell pair and the ECS catch up with the GKP Bell state under bilateral adaptive PNRD around $6.5\,\mathrm{dB}$, corresponding to $\alpha_{\mathrm{cat}}\simeq\alpha_{\mathrm{ECS}}\simeq1.34$. Beyond this point, their CHSH values remain close to, but slightly below, the GKP result. The cat-code Bell-pair and ECS values begin to converge around $\alpha\simeq1.25$ and become nearly indistinguishable at larger amplitudes, as the overlap between the coherent-state components vanishes.}

For the cat-code Bell pair \ozlem{(Fig.~\ref{fig:figure4}(a))}, at $\alpha_{\mathrm{cat}}\simeq0.40$, the bilateral two-stage adaptive receiver gives $S=2.816$, only $0.012$ below the Tsirelson bound and corresponding to $99.6\%$ of $2\sqrt{2}$. Although this is not a like-for-like comparison between identical state resources, it compares favourably with previous ECS proposals: near-maximal values such as $S\simeq2.799$ have been obtained at $\alpha=1.1$ using rotated-parity observables built from a vacuum-selective phase gate~\cite{zhong2019quantum}, \ozlem{whereas the displaced-parity strategy gives only $2.589$ at the same amplitude~\cite{banaszek1998nonlocality, banaszek1999testing} as generalised in Milman \textit{et al.}~\cite{milman2005proposal}.} Here, a comparable and slightly larger violation is obtained with a much smaller cat amplitude using Gaussian preprocessing, photon counting, and two-stage feed-forward.

\section{Discussion}
\label{sec:discussion}
\ozlem{GKP states are widely studied as a bosonic encoding for fault-tolerant quantum computing because their grid structure allows sufficiently small displacement errors to be identified and corrected without changing the encoded logical information. Realising this potential, however, remains experimentally demanding. Fault-tolerant operation is commonly associated with GKP squeezing of approximately $9.75\,\mathrm{dB}$ \cite{larsen2025integrated}, whereas optical demonstrations have so far reached approximately $2.5\,\mathrm{dB}$ \cite{konno2024logical,larsen2025integrated}. This gap motivates the study of quantum-information tasks that can be performed with the more modest squeezing available in current experiments. Here, we have investigated Bell nonlocality in finite-energy GKP Bell states while explicitly accounting for constraints on the local measurements. Our results show that the observable CHSH violation depends strongly on the measurement strategy: although the state supports a violation at low GKP squeezing, homodyne measurements do not always reveal it.}

\ozlem{The unrestricted measurement bound separates this limitation of the measurement from the nonlocality supported by the state. Achieving this bound, however, requires arbitrary dichotomic measurements within the logical subspace. The corresponding measurement directions generally involve setting-dependent logical non-Clifford rotations followed by logical Pauli measurements. A fault-tolerant implementation would therefore require magic-state injection and gate teleportation, with additional imperfections that are not included in the unrestricted bound. This motivates the physical PNRD receivers studied here, which combine Gaussian preprocessing, photon counting and, where required, feed-forward to reveal Bell nonlocality without implementing logical non-Clifford operations.}

\ozlem{Among the receivers considered, bilateral two-stage adaptive PNRD gives the largest best-found CHSH violation for finite-energy GKP Bell states, reaching approximately $95\%$--$98\%$ of the unrestricted value between $0$ and $6\,\mathrm{dB}$. In this regime, the broadened finite-energy structure of the GKP state limits homodyne binning, whereas non-Gaussian detection resolves the relevant correlations more effectively. The adaptive receiver nevertheless requires a tapped mode and conditional feed-forward. A simpler alternative combines optimised homodyne detection for Alice with displaced PNRD for Bob, requiring only a displacement and photon-number-resolving detector on Bob's side. This offers a more accessible route to observing Bell nonlocality with the squeezing already available in optical GKP experiments and, in a loophole-free implementation, could support device-independent applications such as certified randomness.}

\ozlem{The comparison with cat-code Bell pairs and entangled coherent states further shows that the benefit of non-Gaussian detection extends beyond the GKP lattice. The performance of a receiver is instead determined by how well its measurement structure is matched to the bosonic state. Our detector model, which includes finite PNRD efficiency and dark counts, also shows that Bell violations can be maintained for realistic experimental parameters. These results provide concrete receiver targets for experiments with finite-energy bosonic states and demonstrate that strong Bell violations can be obtained without first reaching the squeezing required for fault-tolerant quantum computation or implementing logical non-Clifford control.}

\section*{Acknowledgments}
\label{sec:acknowledgments}
\ozlem{During the preparation of this work, we became aware of the independent work of Salek and Hayashi~\cite{salek2026robust}, which studies robust CHSH self-testing with finite-energy GKP states. Their work is complementary to ours, which focuses on the Bell violations achievable with homodyne and photon-counting-based measurement strategies.}

\ozlem{The research was conceived and directed by the authors. OpenAI's ChatGPT and Codex (GPT-5.6 and GPT-6) were used as assistive tools for implementing and debugging numerical code and refining the manuscript. The authors reviewed the assisted material and take full responsibility for the results and conclusions.}

\section*{Data Availability}
There are no publicly available research data or software supporting this manuscript. Requests for further information or data should be sent to the authors.

\appendix
\section{\label{app:measurement_optimisation}Exact state-support representation and measurement optimisation}
This appendix describes how the physical oscillator states and observables are represented on the exact two-dimensional support of the finite-energy GKP Bell state. We then derive the optimal unrestricted measurements and explain how the same framework is used to optimise the physically constrained homodyne and photon-counting receivers considered in the main text.

\subsection{Löwdin representation of the finite-energy support}
\label{app:lowdin_support}
Let $\ket{k_0}\equiv\ket{\overline{0}_{\Delta}}$ and $\ket{k_1}\equiv\ket{\overline{1}_{\Delta}}$ denote the normalised finite-energy GKP codewords. At finite squeezing, these codewords are generally nonorthogonal. We therefore introduce an orthonormal basis for their two-dimensional support using Löwdin symmetric orthogonalisation~\cite{lowdin1950non,torun2023coherence}. Collecting the codewords into the matrix
\begin{equation}
    K
    =
    \begin{pmatrix}
        \ket{k_0} & \ket{k_1}
    \end{pmatrix},
    \label{eq:app_codeword_matrix}
\end{equation}
whose Gram matrix is
\begin{equation}
    G=K^\dagger K.
    \label{eq:app_gram_matrix}
\end{equation}
The Löwdin-orthogonalised support isometry is then
\begin{equation}
    V_L=KG^{-1/2},
    \qquad
    V_L^\dagger V_L=I_2.
    \label{eq:app_lowdin_isometry}
\end{equation}
The two columns of $V_L$ form an orthonormal basis for the span of the finite-energy codewords. For any physical oscillator operator $O$, its representation on this exact support is
\begin{equation}
    O^{(L)}
    =
    V_L^\dagger O V_L.
    \label{eq:app_operator_compression}
\end{equation}
Equation~\eqref{eq:app_operator_compression} is an exact change of coordinates on the state support. It does not replace the finite-energy codewords by ideal GKP states or assume that physical measurements act as ideal logical Pauli operators.

The pure two-mode state is represented in the truncated Fock basis by a coefficient matrix $M$,
\begin{equation}
    \ket{\Phi_\Delta}
    =
    \sum_{n,m}
    M_{nm}\ket{n}_A\ket{m}_B,
    \qquad
    \operatorname{Tr}(M^\dagger M)=1.
    \label{eq:app_state_coefficient_matrix}
\end{equation}
Because the state is supported on the span of $\ket{k_0}$ and $\ket{k_1}$ on each mode, its coefficient matrix in the Löwdin basis is
\begin{equation}
    M_L
    =
    V_L^\dagger M V_L^*,
    \qquad
    M=V_LM_LV_L^T.
    \label{eq:app_logical_state_matrix}
\end{equation}
For arbitrary local observables $A$ and $B$, expectation values can then be evaluated as
\begin{equation}
    \bra{\Phi_\Delta}A\otimes B\ket{\Phi_\Delta}
    =
    \operatorname{Tr}
    \left[
        M_L^\dagger A^{(L)}M_L
        \left(B^{(L)}\right)^T
    \right].
    \label{eq:app_expectation_identity}
\end{equation}

\subsection{Analytical verification of the unrestricted CHSH bound}
\label{app:overlap_chsh}
\ozlem{The unrestricted CHSH bound can also be obtained directly from the overlap of the normalised finite-energy GKP codewords defined in Eq.~\eqref{eq:gkp_codewords}. This provides an independent check of the numerical Schmidt coefficients and the L\"owdin support representation introduced in the previous section.}

\ozlem{For two Gaussian peaks of equal width, completing the square gives
\begin{align}
    \langle G_\Delta(u)|G_\Delta(v)\rangle
    &=
    \frac{1}{\sqrt{\pi}\Delta}
    \int_{-\infty}^{\infty} dq\,
    e^{-\frac{(q-u)^2+(q-v)^2}{2\Delta^2}}
    \notag\\
    &=
    e^{-(u-v)^2/(4\Delta^2)}.
    \label{eq:overlap_gaussian_peaks}
\end{align}
Writing the codewords as
\begin{equation}
    |\bar{j}_\Delta\rangle
    =
    \frac{1}{\sqrt{N_j}}
    \sum_{s\in\mathbb Z}
    w_{j,s}|G_\Delta(q_{j,s})\rangle,
    \qquad j\in\{0,1\},
\end{equation}
with peak positions $q_{j,s}=(2s+j)\sqrt{\pi}$ and envelope weights $w_{j,s}=\exp[-\Delta^2q_{j,s}^2/2]$, their normalisation constants are
\begin{equation}
    N_j
    =
    \sum_{s,t\in\mathbb Z}
    w_{j,s}w_{j,t}
    \exp\!\left[
        -\frac{(q_{j,s}-q_{j,t})^2}{4\Delta^2}
    \right].
\end{equation}
The mutual codeword overlap is therefore
\begin{align}
    c_\Delta
    &\equiv
    \langle\bar{0}_\Delta|\bar{1}_\Delta\rangle
    \notag\\
    &=
    \frac{1}{\sqrt{N_0N_1}}
    \sum_{s,t\in\mathbb Z}
    w_{0,s}w_{1,t}
    \exp\!\left[
        -\frac{(q_{0,s}-q_{1,t})^2}{4\Delta^2}
    \right].
    \label{eq:overlap_gkp_codewords}
\end{align}
These Gaussian-weighted sums converge absolutely for $\Delta>0$ and include overlaps between distinct peaks. For the real, positive codeword wavefunctions considered here, $c_\Delta$ is real and satisfies $0<c_\Delta<1$.}

\ozlem{The finite-energy Bell state is
\begin{equation}
    |\Phi_\Delta\rangle
    =
    \frac{
        |\bar{0}_\Delta,\bar{0}_\Delta\rangle
        +
        |\bar{1}_\Delta,\bar{1}_\Delta\rangle
    }{\sqrt{\mathcal N}},
    \qquad
    \mathcal N=2(1+c_\Delta^2).
    \label{eq:overlap_bell_state}
\end{equation}
An orthonormal basis for each mode is given by
\begin{equation}
    |u_\pm\rangle
    =
    \frac{
        |\bar{0}_\Delta\rangle
        \pm|\bar{1}_\Delta\rangle
    }{\sqrt{2(1\pm c_\Delta)}}.
    \label{eq:overlap_schmidt_basis}
\end{equation}
Inverting these relations yields
\begin{align}
    |\bar{0}_\Delta\rangle
    &=
    \sqrt{\frac{1+c_\Delta}{2}}\,|u_+\rangle
    +
    \sqrt{\frac{1-c_\Delta}{2}}\,|u_-\rangle,
    \notag\\
    |\bar{1}_\Delta\rangle
    &=
    \sqrt{\frac{1+c_\Delta}{2}}\,|u_+\rangle
    -
    \sqrt{\frac{1-c_\Delta}{2}}\,|u_-\rangle.
\end{align}
Substituting into Eq.~\eqref{eq:overlap_bell_state}, the terms proportional to $|u_+,u_-\rangle$ and $|u_-,u_+\rangle$ cancel, giving the Schmidt decomposition
\begin{equation}
    |\Phi_\Delta\rangle
    =
    \lambda_+|u_+,u_+\rangle
    +
    \lambda_-|u_-,u_-\rangle,
\end{equation}
where
\begin{equation}
    \lambda_\pm
    =
    \frac{1\pm c_\Delta}{\sqrt{2(1+c_\Delta^2)}}.
    \label{eq:overlap_schmidt_coefficients}
\end{equation}
These coefficients are nonnegative and satisfy $\lambda_+^2+\lambda_-^2=1$. The concurrence is consequently
\begin{equation}
    \mathcal C_\Delta
    =
    2\lambda_+\lambda_-
    =
    \frac{1-c_\Delta^2}{1+c_\Delta^2}.
    \label{eq:overlap_concurrence}
\end{equation}
}

\ozlem{Let $X_S$, $Y_S$, and $Z_S$ denote the Pauli operators
in the Schmidt basis $\{|u_+\rangle,|u_-\rangle\}$, with
\begin{align}
    X_S|u_\pm\rangle &= |u_\mp\rangle,
    \notag\\
    Y_S|u_\pm\rangle &= \pm i|u_\mp\rangle,
    \notag\\
    Z_S|u_\pm\rangle &= \pm|u_\pm\rangle.
\end{align}
Here, expectation values are taken in $|\Phi_\Delta\rangle$. The operator $Z_S\otimes Z_S$ leaves both Schmidt components
unchanged, while $X_S\otimes X_S$ exchanges them and $Y_S\otimes Y_S$ exchanges them with a minus sign.}

\ozlem{Operators containing one $Z_S$ and one $X_S$ or $Y_S$ map the state into the span of $|u_+,u_-\rangle$ and $|u_-,u_+\rangle$, which is orthogonal to $|\Phi_\Delta\rangle$. The two remaining cross-axis correlations cancel because the Schmidt coefficients are real
\begin{align}
    \langle X_S\otimes Y_S\rangle
    &=
    \langle Y_S\otimes X_S\rangle
    \notag\\
    &=
    i\lambda_-\lambda_+
    -
    i\lambda_+\lambda_-
    =
    0.
\end{align}
The nonzero two-party Pauli correlations are therefore
\begin{align}
    \langle X_S\otimes X_S\rangle &= \mathcal C_\Delta,
    \notag\\
    \langle Y_S\otimes Y_S\rangle &= -\mathcal C_\Delta,
    \notag\\
    \langle Z_S\otimes Z_S\rangle &= 1.
    \label{eq:overlap_pauli_correlations}
\end{align}
Defining the correlation matrix by
\begin{equation}
    T_{jk}
    =
    \langle\sigma_j\otimes\sigma_k\rangle,
    \qquad j,k\in\{x,y,z\},
\end{equation}
where $(\sigma_x,\sigma_y,\sigma_z)=(X_S,Y_S,Z_S)$, we obtain
\begin{equation}
    T=
    \begin{pmatrix}
        \mathcal C_\Delta & 0 & 0\\
        0 & -\mathcal C_\Delta & 0\\
        0 & 0 & 1
    \end{pmatrix},
\end{equation}
and hence
\begin{equation}
    T^{\mathsf T}T=
    \begin{pmatrix}
        \mathcal C_\Delta^2 & 0 & 0\\
        0 & \mathcal C_\Delta^2 & 0\\
        0 & 0 & 1
    \end{pmatrix}.
\end{equation}
Since $0\leq\mathcal C_\Delta=2\lambda_+\lambda_-\leq\lambda_+^2+\lambda_-^2=1$, the two largest eigenvalues are $\mu_1=1$ and $\mu_2=\mathcal C_\Delta^2$. The Horodecki criterion~\cite{horodecki1995violating} then gives the maximum over both parties' local dichotomic measurements
\begin{align}
    S_{\max}(\Delta)
    &=
    2\sqrt{\mu_1+\mu_2}
    =
    2\sqrt{1+\mathcal C_\Delta^2}
    \notag\\
    &=
    2\sqrt{
        1+
        \left(
            \frac{1-c_\Delta^2}{1+c_\Delta^2}
        \right)^2
    }.
    \label{eq:overlap_chsh_bound}
\end{align}
This recovers the Schmidt-based expression in the main text.}

\ozlem{To show explicitly that the bound is attainable, choose Alice's observables as $A_0=Z_S$ and $A_1=X_S$, and Bob's observables as
\begin{align}
    B_0(\theta)
    &=
    \cos\theta\,Z_S+\sin\theta\,X_S,
    \notag\\
    B_1(\theta)
    &=
    \cos\theta\,Z_S-\sin\theta\,X_S,
\end{align}
with $0\leq\theta\leq\pi/2$. These are valid dichotomic observables because $X_S^2=Z_S^2=I_2$ and $X_SZ_S+Z_SX_S=0$, giving
$B_y(\theta)^2=I_2$. Using Eq.~\eqref{eq:overlap_pauli_correlations}, the CHSH value is
\begin{align}
    S(\theta)
    &=
    \langle A_0\otimes[B_0(\theta)+B_1(\theta)]\rangle
    \notag\\
    &\quad+
    \langle A_1\otimes[B_0(\theta)-B_1(\theta)]\rangle
    \notag\\
    &=
    2\bigl(\cos\theta+\mathcal C_\Delta\sin\theta\bigr).
\end{align}
The maximum occurs at $\tan\theta_{\mathrm{opt}}=\mathcal C_\Delta$, corresponding to
\begin{align}
    B_0
    &=
    \frac{Z_S+\mathcal C_\Delta X_S}
         {\sqrt{1+\mathcal C_\Delta^2}},
    \notag\\
    B_1
    &=
    \frac{Z_S-\mathcal C_\Delta X_S}
         {\sqrt{1+\mathcal C_\Delta^2}}.
\end{align}
These observables give $S(\theta_{\mathrm{opt}})=2\sqrt{1+\mathcal C_\Delta^2}$, saturating Eq.~\eqref{eq:overlap_chsh_bound}. In the orthogonal-codeword limit $c_\Delta\to0$, $\mathcal C_\Delta\to1$ and
$S_{\max}\to2\sqrt{2}$. Conversely, as $c_\Delta\to1$, the concurrence vanishes and $S_{\max}\to2$.
The Pauli operators used here act on the exact state support; their implementation by a particular physical receiver is a separate measurement constraint.}

\ozlem{Finally, the overlap also determines the L\"owdin coefficient matrix. Using the definitions of $K$, $G$, and $V_L$ introduced above,
\begin{equation}
    G=
    \begin{pmatrix}
        1 & c_\Delta\\
        c_\Delta & 1
    \end{pmatrix},
    \qquad
    M=\frac{KK^{\mathsf T}}{\sqrt{\mathcal N}}.
\end{equation}
Since $V_L^\dagger K=G^{1/2}$, the coefficient matrix in the L\"owdin basis is
\begin{align}
    M_L
    &=
    V_L^\dagger M V_L^*
    =
    \frac{
        G^{1/2}(G^{1/2})^{\mathsf T}
    }{\sqrt{\mathcal N}}
    \notag\\
    &=
    \frac{1}{\sqrt{2(1+c_\Delta^2)}}
    \begin{pmatrix}
        1 & c_\Delta\\
        c_\Delta & 1
    \end{pmatrix},
    \label{eq:overlap_lowdin_matrix}
\end{align}
where the last equality follows because $G$ and its positive square root are real symmetric. The singular values of $M_L$ are precisely the Schmidt coefficients in Eq.~\eqref{eq:overlap_schmidt_coefficients}, confirming the consistency of the two representations.}

\subsection{Bob's optimal unrestricted measurement}
\label{app:bob_best_response}
Suppose Alice's observables $A_0$ and $A_1$ are fixed. Their support representations are
\begin{equation}
    A_x^{(L)}
    =
    V_L^\dagger A_xV_L,
    \qquad x\in\{0,1\}.
    \label{eq:app_alice_support_observables}
\end{equation}
Using Eq.~\eqref{eq:app_expectation_identity}, the two CHSH correlation operators acting on Bob's support are
\begin{align}
    C_0^{(L)}
    &=
    \left[
        M_L^\dagger
        \left(A_0^{(L)}+A_1^{(L)}\right)
        M_L
    \right]^T,
    \nonumber\\
    C_1^{(L)}
    &=
    \left[
        M_L^\dagger
        \left(A_0^{(L)}-A_1^{(L)}\right)
        M_L
    \right]^T.
    \label{eq:app_bob_correlation_operators}
\end{align}
The CHSH parameter can therefore be written as
\begin{equation}
    S
    =
    \operatorname{Tr}
    \left[
        C_0^{(L)}B_0^{(L)}
    \right]
    +
    \operatorname{Tr}
    \left[
        C_1^{(L)}B_1^{(L)}
    \right].
    \label{eq:app_chsh_support_form}
\end{equation}

A general binary measurement is represented by an observable satisfying
\begin{equation}
    -I_2\leq B_y^{(L)}\leq I_2.
    \label{eq:app_binary_observable_constraint}
\end{equation}
For a fixed Hermitian operator $C_y^{(L)}$,
\begin{equation}
    \max_{-I_2\leq B_y^{(L)}\leq I_2}
    \operatorname{Tr}
    \left[
        C_y^{(L)}B_y^{(L)}
    \right]
    =
    \left\|C_y^{(L)}\right\|_1,
    \label{eq:app_trace_norm_maximum}
\end{equation}
where $\|\cdot\|_1$ denotes the trace norm. The maximum is attained by the spectral-sign observable. Explicitly, if
\begin{equation}
    C_y^{(L)}
    =
    U_y\Lambda_yU_y^\dagger,
    \label{eq:app_correlation_diagonalisation}
\end{equation}
then Bob's optimal unrestricted observable is
\begin{equation}
    B_y^{\mathrm{opt}}
    =
    \operatorname{sign}
    \left(C_y^{(L)}\right)
    =
    U_y\operatorname{sign}(\Lambda_y)U_y^\dagger.
    \label{eq:app_bob_spectral_sign}
\end{equation}
The value assigned to a zero eigenvalue is arbitrary because the corresponding eigenspace does not contribute to the CHSH value. Bob's optimal contribution for fixed Alice observables is consequently
\begin{equation}
    S_{\mathrm{Bob\,opt}}
    =
    \left\|C_0^{(L)}\right\|_1
    +
    \left\|C_1^{(L)}\right\|_1.
    \label{eq:app_bob_optimal_value}
\end{equation}

The associated two-outcome POVM elements are
\begin{equation}
    M_{\pm|y}
    =
    \frac{I_2\pm B_y^{\mathrm{opt}}}{2}.
    \label{eq:app_optimal_binary_povm}
\end{equation}

\subsection{Alice's optimal response and bilateral optimisation}
\label{app:bilateral_optimisation}
The same construction applies when Bob's observables are fixed and Alice is optimised. In this case, the correlation operators acting on Alice's support are
\begin{align}
    D_0^{(L)}
    &=
    M_L
    \left(
        B_0^{(L)}+B_1^{(L)}
    \right)^T
    M_L^\dagger,
    \nonumber\\
    D_1^{(L)}
    &=
    M_L
    \left(
        B_0^{(L)}-B_1^{(L)}
    \right)^T
    M_L^\dagger.
    \label{eq:app_alice_correlation_operators}
\end{align}
Alice's unrestricted best responses are therefore
\begin{equation}
    A_x^{\mathrm{opt}}
    =
    \operatorname{sign}
    \left(D_x^{(L)}\right),
    \qquad x\in\{0,1\}.
    \label{eq:app_alice_spectral_sign}
\end{equation}

When all four measurements are unrestricted, we alternate between the following updates:

\begin{enumerate}
    \item Fix $A_0^{(L)}$ and $A_1^{(L)}$, construct $C_0^{(L)}$ and $C_1^{(L)}$, and update Bob using Eq.~\eqref{eq:app_bob_spectral_sign}.
    \item Fix the resulting $B_0^{(L)}$ and $B_1^{(L)}$, construct $D_0^{(L)}$ and $D_1^{(L)}$, and update Alice using Eq.~\eqref{eq:app_alice_spectral_sign}.
    \item Repeat the two updates until the change in the CHSH value falls below the chosen convergence tolerance.
\end{enumerate}

Each update is a best response and therefore cannot decrease the CHSH value. For the pure states considered here, the unrestricted result can be checked independently using the Schmidt decomposition
\begin{equation}
    \ket{\Psi}
    =
    \lambda_0\ket{0,0}
    +
    \lambda_1\ket{1,1},
    \qquad
    \lambda_0^2+\lambda_1^2=1.
    \label{eq:app_schmidt_decomposition}
\end{equation}
Defining the concurrence
\begin{equation}
    \mathcal{C}=2\lambda_0\lambda_1,
    \label{eq:app_concurrence}
\end{equation}
the maximum CHSH value is
\begin{equation}
    S_{\max}
    =
    2\sqrt{1+\mathcal{C}^2}.
    \label{eq:app_schmidt_chsh_ceiling}
\end{equation}
Agreement between Eq.~\eqref{eq:app_schmidt_chsh_ceiling} and the alternating optimisation provides an independent validation of the unrestricted numerical result.

\subsection{Optimisation of physically constrained measurements}
\label{app:constrained_measurements}
The spectral-sign observables determine the maximum CHSH value attainable with arbitrary binary measurements on the exact state support. To evaluate experimentally motivated oscillator measurements, we instead optimise each receiver over the family of POVMs that it can physically implement.

Let $\boldsymbol{\theta}_y$ denote the continuous parameters of Bob's receiver for setting $y$. Depending on the measurement architecture, these parameters may include displacement, squeezing, beamsplitter transmissivity, and conditional feed-forward displacements. Let $E_{y,r}(\boldsymbol{\theta}_y)$ denote the corresponding physical POVM, where $r$ is the complete measurement record. Its representation on the state support is
\begin{equation}
    E_{y,r}^{(L)}(\boldsymbol{\theta}_y)
    =
    V_L^\dagger
    E_{y,r}(\boldsymbol{\theta}_y)
    V_L.
    \label{eq:app_povm_compression}
\end{equation}
For a complete POVM,
\begin{equation}
    \sum_r
    E_{y,r}^{(L)}(\boldsymbol{\theta}_y)
    =
    I_2,
    \label{eq:app_povm_completeness}
\end{equation}
with the sum replaced by an integral for a continuous outcome. Assigning the binary value $s_{y,r}\in\{-1,+1\}$ to each record gives the observable
\begin{equation}
    B_y^{(L)}
    \left(
        \boldsymbol{\theta}_y,\mathbf{s}_y
    \right)
    =
    \sum_r
    s_{y,r}
    E_{y,r}^{(L)}(\boldsymbol{\theta}_y).
    \label{eq:app_physical_binary_observable}
\end{equation}
For fixed continuous parameters, define the CHSH contribution of outcome $r$ by
\begin{equation}
    g_{y,r}(\boldsymbol{\theta}_y)
    =
    \operatorname{Tr}
    \left[
        C_y^{(L)}
        E_{y,r}^{(L)}(\boldsymbol{\theta}_y)
    \right].
    \label{eq:app_outcome_score}
\end{equation}
Bob's contribution can then be written as
\begin{equation}
    \operatorname{Tr}
    \left[
        C_y^{(L)}B_y^{(L)}
    \right]
    =
    \sum_r
    s_{y,r}g_{y,r}.
    \label{eq:app_outcome_sum}
\end{equation}
Because every binary sign can be selected independently, the optimal assignment is
\begin{equation}
    s_{y,r}^{\mathrm{opt}}
    =
    \operatorname{sign}
    \left[
        g_{y,r}(\boldsymbol{\theta}_y)
    \right],
    \label{eq:app_optimal_outcome_sign}
\end{equation}
and hence
\begin{equation}
    \max_{\mathbf{s}_y}
    \operatorname{Tr}
    \left[
        C_y^{(L)}B_y^{(L)}
    \right]
    =
    \sum_r
    \left|
        g_{y,r}(\boldsymbol{\theta}_y)
    \right|.
    \label{eq:app_optimised_outcome_sum}
\end{equation}
The remaining continuous receiver parameters are optimised numerically by maximising the right-hand side of Eq.~\eqref{eq:app_optimised_outcome_sum}.

For homodyne detection, the outcome is continuous, $r=q$, and Eq.~\eqref{eq:app_optimised_outcome_sum} becomes
\begin{equation}
    \max_{f_y}
    \operatorname{Tr}
    \left[
        C_y^{(L)}B_y^{\mathrm{hom}}[f_y]
    \right]
    =
    \int_{-\infty}^{\infty}
    dq\,
    \left|g_y(q)\right|,
    \label{eq:app_homodyne_optimum}
\end{equation}
with the optimal binning rule
\begin{equation}
    f_y^{\mathrm{opt}}(q)
    =
    \operatorname{sign}[g_y(q)].
    \label{eq:app_optimal_homodyne_binning}
\end{equation}

For displaced or Gaussian-preprocessed PNRD, the outcome is $r=n$, so
\begin{equation}
    s_{y,n}^{\mathrm{opt}}
    =
    \operatorname{sign}
    \left[
        g_{y,n}(\boldsymbol{\theta}_y)
    \right].
    \label{eq:app_optimal_pnr_binning}
\end{equation}

For the adaptive receiver, the complete record is $r=(m,n)$, where $m$ and $n$ are the outcomes of the first and second PNR detectors. The feed-forward displacement is restricted to
\begin{equation}
    \beta_y(m)
    =
    \begin{cases}
        \beta_y^{(0)}, & m=0,\\
        \beta_y^{(1)}, & m>0.
    \end{cases}
    \label{eq:app_two_branch_feedforward}
\end{equation}
This two-branch rule constrains only the conditional displacement. The first outcome remains photon-number resolved and the full record $(m,n)$ is used in the final binary assignment,
\begin{equation}
    s_{y,mn}^{\mathrm{opt}}
    =
    \operatorname{sign}
    \left[
        g_{y,mn}(\boldsymbol{\theta}_y)
    \right].
    \label{eq:app_optimal_adaptive_assignment}
\end{equation}

When both parties use constrained physical receivers, the optimisation is performed bilaterally. For fixed Alice observables, Bob's continuous parameters and outcome signs are optimised using Eqs.~\eqref{eq:app_outcome_score}--\eqref{eq:app_optimised_outcome_sum}. Alice is then optimised in the same way using the correlation operators $D_x^{(L)}$ in Eq.~\eqref{eq:app_alice_correlation_operators}. These two steps are alternated until convergence.

Unlike the unrestricted spectral-sign update, optimisation over the continuous receiver parameters is generally nonconvex. We therefore use multiple initial parameter sets together with continuation from neighbouring squeezing values and keep the largest converged CHSH value. The resulting value is an achieved lower bound within the chosen receiver family rather than a proof of the global optimum over that family.

\subsection{Logical Pauli decomposition}
\label{app:logical_pauli_decomposition}
To interpret a physical observable on the exact state support, we decompose its Löwdin representation in the Pauli basis:
\begin{equation}
    O^{(L)}
    =
    c_I I_2
    +
    c_X X_L
    +
    c_Y Y_L
    +
    c_Z Z_L.
    \label{eq:app_pauli_decomposition}
\end{equation}
The coefficients are
\begin{equation}
    c_I
    =
    \frac{1}{2}\operatorname{Tr}[O^{(L)}],
    \;\;
    c_j
    =
    \frac{1}{2}
    \operatorname{Tr}
    \left[
        \sigma_jO^{(L)}
    \right],
    \;\;
    j\in\{X,Y,Z\}.
    \label{eq:app_pauli_coefficients}
\end{equation}
A physical observable projected onto the state support need not be a projective qubit observable. In particular, it can have a nonzero identity component and satisfy
\begin{equation}
    \left(O^{(L)}\right)^2\neq I_2.
    \label{eq:app_nonprojective_compressed_observable}
\end{equation}
The coefficients in Eq.~\eqref{eq:app_pauli_decomposition} should therefore be interpreted as the effective action of the physical measurement on the finite-energy state support, rather than as parameters of an assumed ideal logical-qubit measurement.

%\subsection{Numerical truncation and validation}
%\label{app:numerical_validation}
%The oscillator Hilbert space is truncated to the first $N_{\mathrm{cut}}$ Fock states for numerical evaluation. The theoretical photon-number observables and POVMs are defined using sums from $n=0$ to infinity; the sums are truncated only in the numerical implementation. We verify that the probability omitted above the cutoff is below the chosen tolerance.
%For adaptive PNRD, the first- and second-stage photon-number sums are likewise evaluated with finite numerical cutoffs. The retained probability, omitted first-detector tail, POVM completeness error, and population in the highest retained Fock levels are monitored at every optimisation point.
%The final observables are reevaluated using larger validation cutoffs without reoptimising their parameters or binary assignments. A point is accepted only when the CHSH value and relevant probability tails are stable under this cutoff increase.
%For homodyne detection, the continuum integrals are evaluated over a finite quadrature interval chosen sufficiently large that the omitted probability is negligible. Integrals involving discontinuous binning functions are divided at every bin boundary before numerical quadrature is performed.
%The bilateral iteration is terminated when the increase in the CHSH value falls below the specified convergence tolerance. Multiple parameter initialisations and continuation in the state squeezing or coherent-state amplitude are used to reduce sensitivity to local optima.

\section{Convergence of optimised homodyne binning to periodic GKP binning}
\label{app:binning_strategies_figure}
\begin{figure}[htbp]
%\hspace*{-0.3cm}
\includegraphics[scale=0.44]{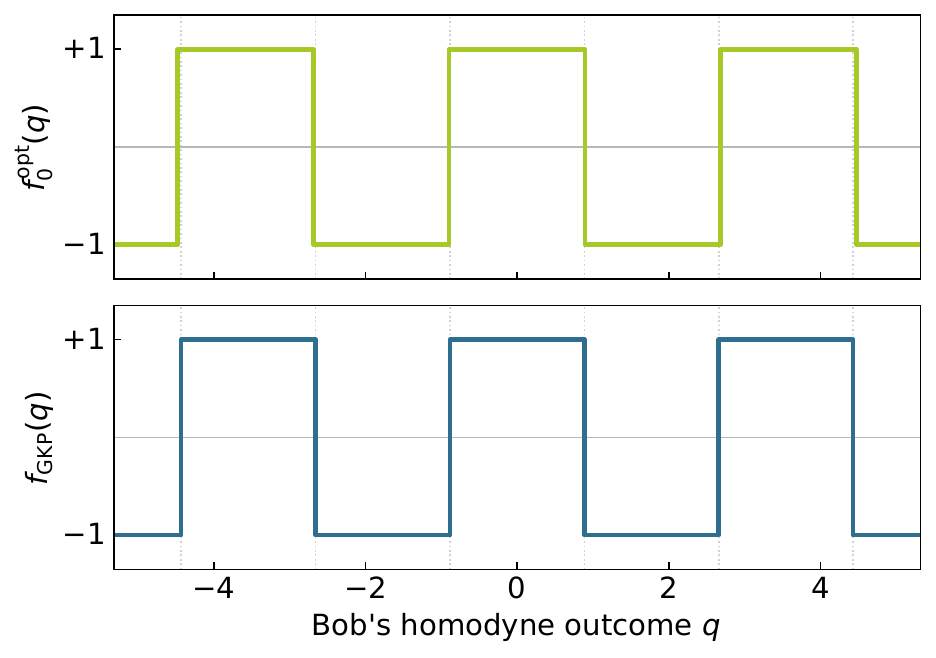}\hspace*{1cm}
\caption{\label{fig:figure1_appendix}Homodyne binning functions underlying Fig.~\ref{fig:figure1}(a) at $10\,\mathrm{dB}$ of GKP squeezing. The upper panel shows Bob's optimised binary filter $f_{0}^{\mathrm{opt}}(q)$ for the $y=0$ CHSH setting, obtained from the joint optimisation of Alice's and Bob's homodyne binning functions. The lower panel shows the standard periodic GKP binning function $f_{\mathrm{GKP}}(q)$. The vertical dotted lines indicate the periodic GKP bin boundaries. At this squeezing level, the optimised filter has converged to the periodic assignment over the displayed range, consistent with the near coincidence of the two CHSH curves in Fig.~\ref{fig:figure1}(a).
}
\end{figure}

In this section, we show that the optimised homodyne binning function converges to the standard periodic GKP binning as the GKP squeezing increases. At finite squeezing, the optimised filter assigns each quadrature outcome according to its contribution to the CHSH value, and therefore differs from the periodic rule when the GKP peaks are broadened and shaped by the finite-energy envelope. Figure~\ref{fig:figure1_appendix} shows the corresponding filters at $10\,\mathrm{dB}$ of GKP squeezing. In this high-squeezing regime, the optimised filter has converged to the periodic GKP binning function over the relevant quadrature range. This explains why the periodic and optimised homodyne-binning curves become nearly indistinguishable above approximately $8\,\mathrm{dB}$ in Fig.~\ref{fig:figure1}(a).

\ozlem{\section{Parameters Used}
\label{app:parameters_used}
\begin{figure}[htbp]
%\hspace*{-0.3cm}
\includegraphics[scale=0.42]{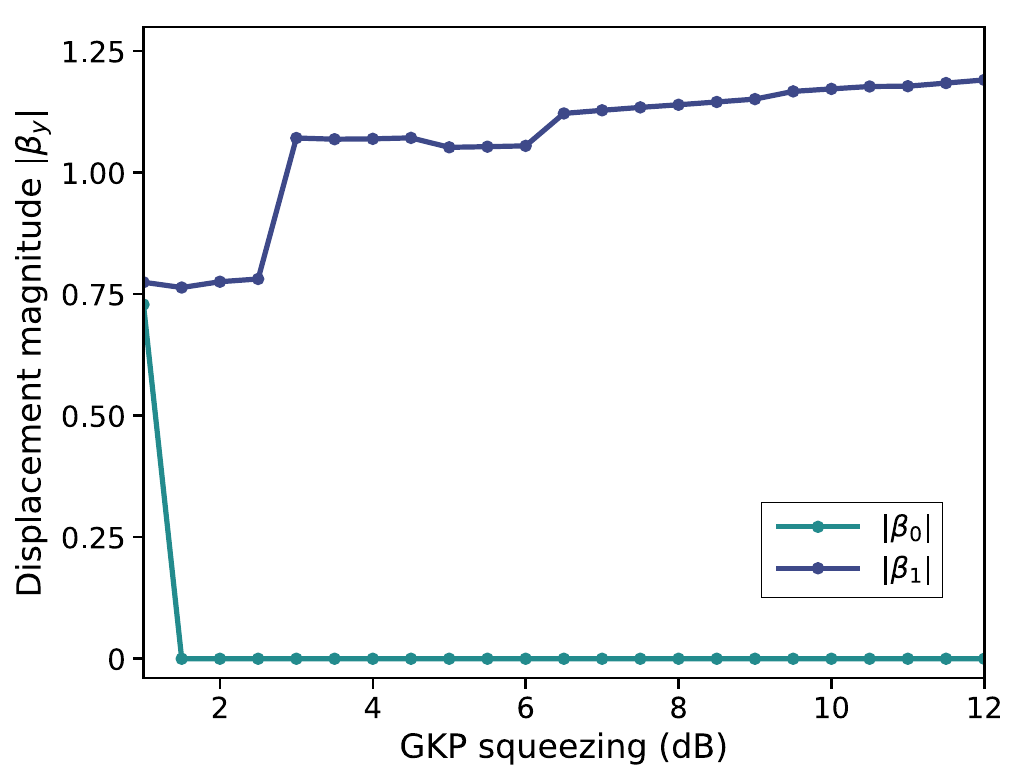}\hspace*{-0.1cm}
\caption{\label{fig:figure2_appendix}Optimised displacement magnitudes for Bob's two settings in the homodyne displaced-PNRD strategy of Fig.~\ref{fig:figure3}. Each marker denotes an independently optimised value at the indicated GKP squeezing. Above approximately $1.5\,\mathrm{dB}$, the $y=0$ setting is implemented without displacement, while the $y=1$ setting retains a finite displacement.
}
\end{figure}
This appendix summarises the numerical settings used throughout the GKP calculations. We represent the finite-energy GKP codewords and Bell state in a Fock basis with $N_{\mathrm{cut}}=200$. The codeword wavefunctions are constructed over $q\in[-20,20]$ on a uniform grid with spacing $\delta q=0.005$, keeping lattice peaks with indices $k=-10,\ldots,10$.}

\ozlem{All homodyne observables are evaluated by numerical quadrature over $q\in[-14.5\sqrt{\pi},14.5\sqrt{\pi}]$, with the quadrature grid chosen so that the periodic-binning boundaries lie at integration-cell boundaries. We use the same grid for the periodic and optimised homodyne-binning calculations. For optimised homodyne binning, we allow the binary assignment $f(q)\in\{-1,+1\}$ to vary across the quadrature grid. We initialise the optimisation with the periodic and constant-output filters, as well as $12$ random filters, and perform at most $80$ see-saw iterations. The unrestricted benchmark is obtained from $12$ random initial measurements with at most $200$ iterations.}

\ozlem{The physical receiver results in Fig.~\ref{fig:figure3} are calculated from $0$ to $12\,\mathrm{dB}$ of GKP squeezing in steps of $0.5\,\mathrm{dB}$. For each squeezing value, we optimise the receiver parameters and photon-number assignments independently for each party and CHSH setting. Although the GKP state itself is represented using $N_{\mathrm{cut}}=200$, the receiver calculations require a larger Fock space because displacement, squeezing and beam-splitter operations populate higher photon-number states. We therefore choose the receiver cutoff separately for each measurement architecture and verify convergence by repeating the calculation at a larger cutoff. For displaced PNRD, we use a Fock cutoff of $400$ and allow $|\operatorname{Re}\beta_y|,|\operatorname{Im}\beta_y|\leq6$. For squeezed-displaced PNRD, we use a cutoff of $520$, the same displacement range, and real squeezing parameters $\zeta_y\in[-1,1]$. The bilateral squeezed-displaced PNRD calculation without a tapped mode uses a cutoff of $700$, as Gaussian preprocessing is applied independently on both modes. For the two-stage adaptive PNRD receiver, we use a Fock cutoff of $520$ and resolve the first photon-number measurement up to $m=239$. The beam-splitter transmissivity is varied over $0\leq T_y\leq1$. The displacement $\alpha_y$ before the first PNRD and the conditional displacements $\beta_y^{(0)}$ and $\beta_y^{(1)}$ are restricted to a single quadrature axis, with amplitudes no larger than $6$. We similarly use $\zeta_y\in[-1,1]$. The two feed-forward branches correspond to $m=0$ and $m\geq1$, while the full record $(m,n)$ is retained when assigning the final binary outcome.}

\ozlem{For the detector-noise calculations, we take a detection window of $\tau_{\mathrm{gate}}=100\,\mathrm{ns}$, so that $\lambda_{\mathrm{dc}}=R_{\mathrm{dc}}\tau_{\mathrm{gate}}$. The bilateral adaptive-PNRD efficiency sweep uses $\eta_d=0.85$, $0.90$, and $0.95$ at $R_{\mathrm{dc}}=50\,\mathrm{Hz}$. For the dark-count sweep, we fix $\eta_d=0.95$ and use $R_{\mathrm{dc}}=50$, $100$, $1000$, and $2000\,\mathrm{Hz}$. In the homodyne-adaptive PNRD calculation, Alice's homodyne detector has efficiency $\eta_{\mathrm{hom}}=0.995$ and $20\,\mathrm{dB}$ electronic-noise clearance. Bob's PNRDs use $R_{\mathrm{dc}}=50\,\mathrm{Hz}$ and the efficiency specified for each curve.}

\begin{figure*}[htbp]
%\hspace*{-0.3cm}
\includegraphics[scale=0.45]{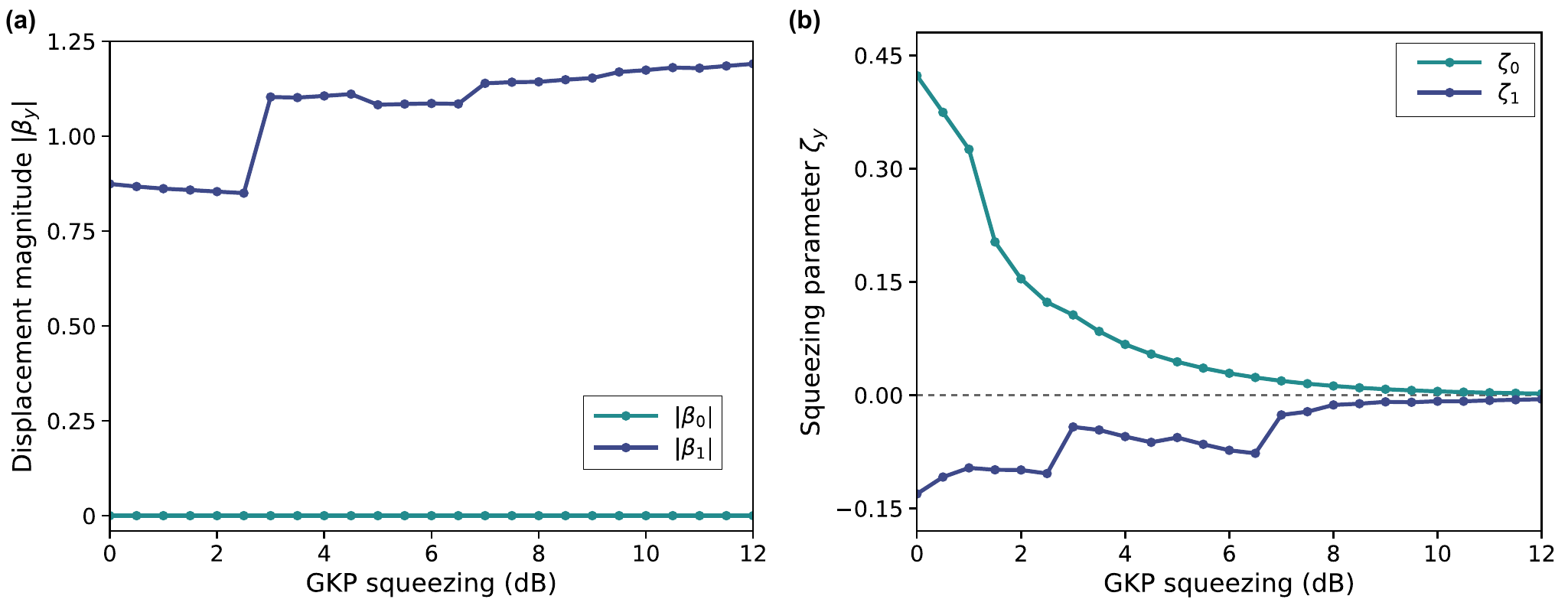}\hspace*{-0.1cm}
\caption{\label{fig:figure3_appendix}
Optimised receiver parameters for the homodyne squeezed-displaced PNRD strategy in Fig.~\ref{fig:figure3}. (a) Bob's displacement magnitudes and (b) real squeezing parameters for the two CHSH settings, $y=0,1$. Each marker denotes an independently optimised value at the indicated GKP squeezing. One setting remains undisplaced, while the squeezing is largest at low GKP squeezing and approaches zero in the high-squeezing regime.
}
\end{figure*}

\ozlem{Finally, Figs.~\ref{fig:figure2_appendix} and \ref{fig:figure3_appendix} show the optimised parameters of the hybrid PNRD receivers. For the homodyne displaced PNRD receiver, Bob's $y=0$ setting becomes an undisplaced PNR measurement above approximately $1.5\,\mathrm{dB}$, whereas the $y=1$ setting uses a finite displacement. The squeezed-displaced receiver follows the same pattern, but additionally uses finite squeezing at low GKP squeezing. Both squeezing parameters decrease towards zero as the GKP squeezing increases, consistent with Gaussian preprocessing providing its main benefit in the finite-squeezing regime.}

\ozlem{\section{Modelling inefficient and noisy PNRDs}
\label{app:noisy_pnrd}
In this section, we describe how finite detection efficiency and dark counts are incorporated into the PNRD model. Following Ref.~\cite{lee2004towards}, each imperfect detector is represented by a classical response matrix. If $n$ photons arrive at a detector, the probability of recording $\tilde{n}$ photons is
\begin{equation}
    Q_{\tilde{n}|n}(\eta_d,\lambda_{\mathrm{dc}})
    =
    e^{-\lambda_{\mathrm{dc}}}
    \sum_{j=0}^{\min(\tilde{n},n)}
    \binom{n}{j}
    \eta_d^j
    (1-\eta_d)^{n-j}
    \frac{
        \lambda_{\mathrm{dc}}^{\tilde{n}-j}
    }{
        (\tilde{n}-j)!
    },
    \label{eq:noisy_pnrd_response}
\end{equation}
where $\eta_d$ is the detection efficiency and $j$ is the number of successfully detected signal photons. The remaining $\tilde{n}-j$ recorded photons arise from dark-count events, which are assumed to follow a Poisson distribution with mean $\lambda_{\mathrm{dc}}$ per detection window. This mean is related to the dark-count rate $R_{\mathrm{dc}}$ and the detection-window duration $\tau_{\mathrm{gate}}$ through
\begin{equation}
    \label{eq:mean_dark_count_number}
    \lambda_{\mathrm{dc}}
    =
    R_{\mathrm{dc}}\tau_{\mathrm{gate}}.
\end{equation}}
\ozlem{Note that for the adaptive receiver, the feed-forward displacement in Eq.~\eqref{eq:adaptive_displacement} is conditioned on the recorded first detector outcome $\tilde m$.}

\ozlem{\section{\label{app:homodyne_efficiency}Modelling homodyne efficiency and electronic noise}
For the hybrid strategies in Fig.~\ref{fig:figure3}(e) and (f), Alice's homodyne detector is modelled using a classical Gaussian response on the measured quadrature. After calibrating the recorded outcome to the input quadrature scale, the probability density of recording $\tilde q$ when the ideal outcome is $q$ is
\begin{equation}
    H(\tilde q|q)
    =
    \frac{1}{\sqrt{2\pi\sigma_{\mathrm{hom}}^2}}
    \exp\!\left[
        -\frac{(\tilde q-q)^2}
        {2\sigma_{\mathrm{hom}}^2}
    \right],
    \label{eq:noisy_homodyne_response}
\end{equation}
where
\begin{equation}
    \sigma_{\mathrm{hom}}^2
    =
    \frac{
        1-\eta_{\mathrm{hom}}+\epsilon_{\mathrm{el}}
    }{
        2\eta_{\mathrm{hom}}
    },
    \qquad
    \epsilon_{\mathrm{el}}
    =
    10^{-C_{\mathrm{el}}/10}.
    \label{eq:noisy_homodyne_variance}
\end{equation}
Here, $\eta_{\mathrm{hom}}$ is the optical homodyne efficiency and $C_{\mathrm{el}}$ is the electronic-noise clearance. Equation~\eqref{eq:noisy_homodyne_variance} uses the quadrature convention of the main text, for which the vacuum variance is $1/2$.}

\ozlem{For a binary binning function $f(\tilde q)\in\{-1,+1\}$ applied to the recorded outcome, the corresponding effective binning of the ideal outcome is
\begin{equation}
    \overline{f}(q)
    =
    \int_{-\infty}^{\infty}
    d\tilde q\,
    H(\tilde q|q)f(\tilde q).
    \label{eq:noisy_homodyne_effective_binning}
\end{equation}
The noisy binned-homodyne observable is therefore
\begin{equation}
    A_{q,\mathrm{noisy}}^{(N)}[f]
    =
    \int_{-\infty}^{\infty}
    dq\,
    \overline{f}(q)
    P_N\ket{q}\!\bra{q}P_N,
    \label{eq:noisy_projected_homodyne}
\end{equation}
with the analogous expression for $p$-homodyne detection.}

\ozlem{Numerically, $H(\tilde q|q)$ is evaluated as a column-normalised response matrix on the same quadrature grid used to construct $A_q^{(N)}[f]$ and $A_p^{(N)}[f]$. The binary filter is then reoptimised over the recorded outcome $\tilde q$, so the homodyne inefficiency and electronic noise are included before the outcome assignment. We use $\eta_{\mathrm{hom}}=0.995$ and $C_{\mathrm{el}}=20\,\mathrm{dB}$, giving $\epsilon_{\mathrm{el}}=0.01$ and $\sigma_{\mathrm{hom}}\simeq0.0868$. For reference, this total noise is equivalent to an effective homodyne efficiency $\eta_{\mathrm{hom}}/(1+\epsilon_{\mathrm{el}})\simeq0.9851$.
}

\section{Energy matching for the state comparison}
\label{app:energy_matching}
In order to make a fair comparison between the finite-energy GKP Bell state, cat-code Bell state, and ECS, we match their mean photon number per mode. We denote the reduced-state energy of the GKP Bell pair by
\begin{equation}
    \overline{n}_{\mathrm{GKP}}(s_{\mathrm{GKP}})
    =
    \bra{\Phi_{\Delta}}
        \hat a_A^{\dagger}\hat a_A
    \ket{\Phi_{\Delta}},
    \label{eq:gkp_energy_per_mode}
\end{equation}
which is identical for Alice and Bob. For the cat-code Bell pair and the ECS,
the corresponding energies are expressed as
\begin{align}
    \overline{n}_{\mathrm{cat}}(\alpha)
    &=
    \frac{\alpha^2}{2}
    \left[
        \tanh(\alpha^2)+\coth(\alpha^2)
    \right]
    =\alpha^2\coth(2\alpha^2),
    \label{eq:cat_energy_per_mode}
    \\
    \overline{n}_{\mathrm{ECS}}(\alpha)
    &=
    \alpha^2
    \frac{1-e^{-4\alpha^2}}{1+e^{-4\alpha^2}}
    =\alpha^2\tanh(2\alpha^2). 
    \label{eq:ecs_energy_per_mode}
\end{align}
For every $s_{\mathrm{GKP}}$, the matched amplitudes
$\alpha_{\mathrm{cat}}$ and $\alpha_{\mathrm{ECS}}$ are determined separately
from
\begin{equation}
    \overline{n}_{\mathrm{cat}}(\alpha_{\mathrm{cat}})
    =
    \overline{n}_{\mathrm{ECS}}(\alpha_{\mathrm{ECS}})
    =
    \overline{n}_{\mathrm{GKP}}(s_{\mathrm{GKP}}).
    \label{eq:energy_matching_condition}
\end{equation}
The two amplitudes are generally different because the ECS is strongly
vacuum-dominated at small $\alpha$, whereas the odd cat codeword approaches a
single-photon state in the same limit.

%\noindent \textbf{Data availability:}
%The datasets generated and/or analysed during the current study are not publicly available because they consist of numerically generated data that can be reproduced from the equations and methods described in this manuscript, but are available from the corresponding author on reasonable request.

%\noindent \textbf{Code availability:}
%The codes that support the findings of this study are available from the corresponding author upon reasonable request.

%\noindent \textbf{Acknowledgments:}
%This research was funded by the Australian Research Council Centre of Excellence for Quantum Computation and Communication Technology (Grant No. CE170100012) and by A*STAR grants C230917010 (Emerging Technology), C230917004 (Quantum Sensing) and Q.InC Strategic Research and Translational Thrust.
%\vspace{0.3cm}

%\noindent \textbf{Author contributions:}
%\ozlem{SMA conceived the project and proposed the use of an optical parametric amplifier. OE developed the full protocol, including the methods for generating GKP states and other non-Gaussian states from the OPA-based architecture, with input from AD, BS and TCR. PKL, TCR and SMA supervised the project. OE performed the simulations and drafted the manuscript with contributions from all authors.}

%\noindent \textbf{Competing Interests:} Authors Ping Koy Lam and Timothy C. Ralph are Editorial Board Members of npj Quantum Information. They were not involved in the journal’s review of, or decisions related to, this manuscript. The other authors declare no other competing interests.

%\section*{References}
\bibliography{apssamp}% Produces the bibliography via BibTeX.

\bibliographystyle{naturemag}

\clearpage

\end{document}